\documentclass[12pt]{iopart}

\usepackage{bm}%¼Ó´Ö×ÖÄ¸
\usepackage{csquotes}%ÅÅ°æÒýºÅ
 \usepackage{amsfonts}
 \usepackage{mathrsfs}
 \usepackage{amsfonts}
 \usepackage{color,xcolor}
 \usepackage[ruled]{algorithm2e}
\usepackage{graphicx}
\usepackage{subfig}
\usepackage{graphicx}
\usepackage{exscale}
\usepackage{color}
\usepackage{times}
\usepackage{epsfig}
\usepackage{cite}
\usepackage[colorlinks,linkcolor=red]{hyperref}
\makeatletter
\@addtoreset{equation}{section}
\makeatother
\renewcommand{\theequation}{\arabic{section}.\arabic{equation}}

\usepackage[abs]{overpic}  % includegraphics with a overprinted tag
\begin{document}

\title[Edge adaptive hybrid regularization model for image deblurring]{Edge adaptive hybrid regularization model for image deblurring}

\author{Tingting Zhang$^1$, Jie Chen$^1$, Caiying Wu$^1$,  Zhifei He$^1$,  Tieyong Zeng$^2$ and Qiyu Jin$^1$}

\address{$^1$ School of Mathematical Science, Inner Mongolia University, Hohhot, China\\
         $^2$ Department of Mathematics, The Chinese University of Hong Kong, Shatin, Hong Kong, China\\
}
\eads{\mailto{ztt2019@aliyun.com}, \mailto{jchenimu@aliyun.com}, \mailto{Haraph@aliyun.com}, \mailto{wucaiyingyun@163.com}, \mailto{zeng@math.cuhk.edu.hk} and \mailto{qyjin2015@aliyun.com}}
\vspace{10pt}
%\begin{indented}
%\item[]December 2020
%\end{indented}

\begin{abstract}
% The  parameter  selection  is  crucial  to  regularization  based  image  restorationmethods.   Generally  speaking,  a  spatially  fixed  parameter  of  regularization  item  for  wholeimages  doesn’t  perform  well  both  at  edges  and  smooth  areas.A  large  parameter  ofregularization item reduces noise better in smooth area but blurs edges, while a small parametersharpens  edges  but  causes  residual  noise.   In  this  paper,  an  automated  spatially  dependentregularization parameter hybrid regularization model is proposed for reconstruction of noisyand  blurred  images  which  combines  the  harmonic  and  TV  models.   The  algorithm  detectsimage edges and spatially adjusts the parameters of Tikhonov and TV regularization termsfor each pixel according to edge information.  In addition, the edge information matrix willbe dynamically updated with the iteration process.  Computationally,  the newly-establishedmodel  is  convex,  then  it  can  be  solved  by  the  semi-proximal  alternating  direction  methodof  multipliers  (sPADMM)  with  a  linear-rate  convergence.Numerical  simulation  resultsdemonstrate that the proposed model effectively protects the image edge while eliminatingnoise and blur and outperforms the state-of-the-art algorithms in terms of PSNR, SSIM andvisual quality.

 The parameter selection is crucial to regularization based image restoration methods.
Generally speaking, a spatially fixed parameter for regularization item in  the whole image does not perform well for both edge and smooth areas. A larger parameter of regularization item reduces noise better in smooth areas but blurs edge regions,  while a small parameter sharpens edge but causes residual noise. 
In this paper, an automated spatially adaptive regularization model, which combines the harmonic and TV models, is proposed for reconstruction of noisy and blurred images. 
In the proposed model, it detects the edges  and then spatially adjusts the parameters of Tikhonov and TV regularization terms for each pixel according to the edge information. 
Accordingly,  the edge information matrix will be also  dynamically updated during the iterations.  
Computationally, the newly-established model is convex, which can be solved by  the semi-proximal alternating direction method of multipliers (sPADMM) with a linear-rate convergence rate. Numerical simulation results demonstrate that the proposed model effectively reserves the image edges and eliminates the noise and blur at the same time. 
In comparison to  state-of-the-art algorithms, it outperforms other methods in terms of PSNR,  SSIM and visual quality.
\end{abstract}

\submitto{\IP}
%\maketitle

\section{Introduction}

Images are frequently degraded by noise and blurring during the image acquisition and transmission procedures ~\cite{2013N, 2014Al}.
As a result, image denoising and deblurring are two fundamental steps for various image processing tasks,  such as image segmentation,  edge detection,  and pattern recognition.
Let $\mathbf{u}\in \mathbb{R}^{M\times N}$ be the unknown clear image of size $M\times N$,  and $\mathbf{f}\in \mathbb{R}^{M\times N}$ be the observed degraded image with Gaussian white noise. Then the mathematical model for image degradation ~\cite{1994Analysis, 2008The, 2014Image, 2016An} is formulated as follows,
\begin{equation}
\mathbf{f}
= \mathbf{A\otimes u} + \bm{\varepsilon},
\end{equation}
where $\mathbf{A}$ is the known blurring operator, $\mathbf{A\otimes u}$ is  the convolution of $\mathbf{A}$ with $\mathbf{u}$ as
\begin{equation*}
[\mathbf{A}\otimes
%
%\nabla
\mathbf{u}](i, j) =  \sum_{(s, t)\in \Omega} \mathbf{A}[(i, j)-(s, t)]\times %\nabla
\mathbf{u}(s, t).
\end{equation*}
 Moreover, $\bm{\varepsilon}\in  \mathbb{R}^{M\times N}$ represents the Gaussian white noise with mean $0$ and standard deviation $\sigma$.
Let $\mathcal{F}$ denotes the vector space of 2D images defined on
$\Omega =\{1, 2, \cdots, M\}\times \{ 1, 2, \cdots, N \}$.
For each 2D image $\mathbf{u}$,  the total number of pixels is $M\times N$,  and $\mathbf{u}(i, j)$ denotes the image value at pixel $(i, j)\in \Omega$.
Restoring the unknown image $\mathbf{u}$ from the degraded image $\mathbf{f}$  is a typical ill-posed problem.
Hence,  effective image deblurring methods usually rely on delicately designed regularizations based on the prior information on $\mathbf{u}$.

To preserve significant edges,  Rudin,  Osher,  and Fatemi\cite{1992Nonlinear} proposed the celebrated total variation (TV) model for image restoration.
By this approach,  the recovery of the image $\mathbf{u}$ is based on solving the following minimization problem
\begin{equation}
\min _{\mathbf{u}}
\frac{\mu}{2}\|\mathbf{A\otimes u-f}\|_{2}^{2}+
\|\nabla\mathbf{u}\|_{1},
\label{Model: rof}
\end{equation}
where  $\nabla \mathbf{u}=[\nabla_{x}\mathbf{u}, \nabla_{y}\mathbf{u}] \in \mathbb{R}^{2M\times N} $ is the gradient in the discrete setting
with $\nabla \mathbf{u} (i, j)=(\nabla_{x} \mathbf{u} (i, j), \nabla_{y} \mathbf{u} (i, j))$,  and $\nabla_{x} \mathbf{u}$ and $\nabla_{y} \mathbf{u}$ denote the horizontal and vertical first-order differences with Dirichlet boundary condition respectively,  which are defined as follows
\begin{equation}\label{diff x}
\eqalign{\nabla_{x}\mathbf{u}(i, j)=\cases{\mathbf{u}(i+1, j)-\mathbf{u}(i, j), &if \quad $i < M;$\\
0, &if \quad $i = M,$\\} \cr
\nabla_{y}\mathbf{u}(i, j)=\cases{\mathbf{u}(i, j+1)-\mathbf{u}(i, j), &if \quad $j < M;$\\
0, &if \quad $j = M,$\\}}
\end{equation}
and
$\|\nabla\mathbf{u}\|_{1}$ is the $l_1$-norm of $\nabla  \mathbf{u} $
\begin{equation}
\qquad \quad \fl  \| \nabla\mathbf{u}\|_{1} = \sum_{(i, j)\in \Omega} |\nabla \mathbf{u} (i, j)|, \quad
|\nabla \mathbf{u} (i, j)|=\sqrt{ (\nabla_{x} \mathbf{u})^{2}(i, j)+ (\nabla_{y}\mathbf{u})^{2}(i, j)}.
\label{norm tv}
\end{equation}
As the total variation term is powerful for preserving sharp edges,  the TV model has been widely used for image processing.
Over the past years,  much effort has been devoted to studying,  solving and extending the TV model.
In particular,  a two-phase approach for deblurring images ~\cite{2008Two, 2010Fast} was proposed to handle the impulsive noise. More related models and algorithms have been reported in ~\cite{2009Linearized, 2009Split, 2004An, 2013Fractional, 2015An, 2009The, 2008Af, 2017Low, 2018Simultaneous, 2013Non, 1990Scale, 2008Anew, 2010Augmented, 2020Ac, 2017Nonconvex, 2012Av}.

However,  image deblurring with TV model often leads to some undesirable staircase effects,  namely,  the transformation of smooth regions into piece-wise constant ones.
There are at least two possible ways to handle this issue.
One is to use the tight-frame approaches ~\cite{2009Linearized, 2009Split}.
The other is to combine the TV regularization term with some more advanced models.
For instance, the harmonic model ~\cite{1977Solution} with Tikhonov regularization and fourth-order PDE (LLT) filter ~\cite{2003Noise} can effectively reduce noise.
Therefore,  some hybrid regularization models combining the TV model with LLT models ~\cite{2007Image, 2016Hybrid, 2003Noise, 2013Non} or harmonic models ~\cite{2013A, 2014Ana} are proposed to preserve edges while restraining the staircase effects at the same time.
Furthermore,  Liu et al.\cite{2020Ani} combines image sharpening operator and framelet regularization ~\cite{2012Image} for image deblurring,  whose model is expressed as,
\begin{equation}
% \min\limites_{\mathbf{u}}\frac{\mu}{2}\|\mathbf{Au-f}\|_{2}^{2}+\|\mathbf{Wu}\|_{1}+\alpha\|\mathbf{u}-B(\mathbf{f})\|_{1},
\min_{\mathbf{u}}\frac{\mu}{2}\|\mathbf{A\otimes u-f}\|_{2}^{2}+\|\mathbf{Wu}\|_{1}+\alpha\|\mathbf{u}-B(\mathbf{f})\|_{1},
\end{equation}
where $\mathbf{W}$ is the framelet transformation,  $B(\mathbf{f})\in  \mathbf{R}^{M\times N} $ is a sharpened image,  and $\mu,  \alpha$ are positive parameters.
In the optimization problem (\ref{Model: rof}),  the positive parameter $\mu$ controls the trade-off between a good fit of $\mathbf{f}$ and a smoothness requirement by the total variation regularization.
In general,  an image is  composed of multiple objects at different scales.
This suggests that different values of $\mu$  are desirable for various image features of different scales, to obtain better results.
More specifically,  for the texture regions,  we can use larger $\mu$ as it leads to less smoothed and more detailed restoration.
On the other hand,  for the flat region,  smaller $\mu$ is desired  to get a better smoothing and noise-reduced result.
For this reason,  in ~\cite{2008A, 2003TV} the multi-scale total variation (MTV) models were proposed,  with spatially varying parameters based on (\ref{Model: rof}).
The corresponding multi-scale version of model (\ref{Model: rof}) is represented as
\begin{equation}
\min_{\mathbf{u}}\frac{1}{2}  \|\mathbf{A \otimes u-f}\|_{\bm{\mu}}^{2}+\| \nabla\mathbf{u}\|_{1},
\label{Model: MTVf}
\end{equation}
where $\bm{\mu}\in  \mathbb{R}^{M\times N}$ is spatially varying non-negative parameter matrix,
$\|\mathbf{A\otimes u-f}\|_{2, \bm{\mu}}^{2}={\langle\mathbf{A\otimes u-f},  \bm{\mu}.*(\mathbf{A\otimes u-f})\rangle}$.
As in Matlab,  $\bm{\mu}.*(\mathbf{A\otimes u-f})$ denotes point-wise product between the elements $\bm{\mu} $ and $\mathbf{A\otimes u-f}$ as follows,
\begin{equation}
[\bm{\mu}.*(\mathbf{A\otimes u-f})](i, j) =\bm{\mu}(i, j)\times [\mathbf{A\otimes u-f}](i, j).
\label{product pw}
\end{equation}
Borrowing the idea of (\ref{Model: MTVf}),  the multi-scale technique can be applied to the TV term,  which yields
\begin{equation}
% \min\limites_{\mathbf{u}}
\min_{\mathbf{u}}
\frac{\mu}{2}\|\mathbf{A\otimes u-f}\|_{2}^{2}+
\| \nabla \mathbf{u} \|_{1, \bm{\alpha}},
\label{Model: MTV2}
\end{equation}
where $\bm{\alpha} \in  \mathbb{R}^{M\times N} $ is a non-negative spatially varying parameter matrix,  $\| \nabla \mathbf{u} \|_{1, \bm{\alpha}}$ is given by
\begin{equation}
\| \nabla\mathbf{u}\|_{1, \bm{\alpha}} =  \sum_{(i, j)\in \Omega}\bm{\alpha}(i, j) |\nabla \mathbf{u} (i, j)|.%\quad |\nabla \mathbf{u} (i, j)|=\sqrt{ \mathbf{u}_{x}^{2}(i, j)+ \mathbf{u}_{y}^{2}(i, j)}
\label{norm 1 alpha}
\end{equation}
Introducing a local parameter gives more flexibility to the model in exchange for the robustness.
For this reason, Dong et al.\cite{2011Automated} proposed a method to decide whether to use a spatially varying parameter,  by using a confidence interval technique based on the expected maximal local variance estimate.
In this paper,  we study the properties of the image edges and propose an edge adaptive hybrid regularization model by introducing an edge detection operation to a combination of the TV and harmonic models.
The edge detection operation is used to decide,  in a robust way,  where the edges are located in the noisy image $\mathbf{f}$ and generates an edge information matrix. This edge information matrix, is later applied to decide  the acceptance or rejection of a scaling parameter. It is worth to mention that
our edge information matrix is obtained and updated fully automatically in each iteration for preserving edges.
Furthermore,  the proposed model is convex once the local parameters are fixed,  and hence it can be solved efficiently by the semi-proximal alternating direction method of multipliers (sPADMM) \cite{fazel2013hankel,li2016majorized,han2018linear},
which  has a linear convergence rate under some mild conditions. The contributions of this work are summarized as follows:
\begin{enumerate}
  \item we propose an algorithm which can automatically detect the image edges and adjust the parameters for the Tikhonov and TV regularization terms for each pixel according to the  edge information.  This enables the proposed algorithm to effectively remove noise on the smooth area and sharpen the edges during deblurring;
  %We combine the harmonic model with the total variation for image deblurring,
  %which is better than applying either one alone. The key idea is that the edge adaptive parameters for TV and Tikhonov regularization terms can treat the edges and the smooth areas differently,  which improves the denoising performance reasonably.
  \item we build a convex optimization model which is easily solved by sPADMM with a linear-rate convergence rate;
  \item we conduct extensive experiments to prove that the proposed method outperforms the state-of-the-art methods in restoring the noisy and blurred images.
  %Our algorithm (EAHR) has competitive performance in sharpening the edges and removing the noise by comparing it with related state-of-the-art methods.
\end{enumerate}
The remainder of the paper is organized as follows.
% In section 2,  we propose our model by adjusting the parameters of TV and Tikhonov regularization terms according to edge information.
In Section 2,  we present our model which describes how to adjust the parameters of TV and Tikhonov regularization terms according to edge information.
The simulation results are exhibited and analyzed in Section 3,  followed by concluding remarks in Section 4.

\section{Edge adaptive hybrid regularization model for image deblurring}

In this section,  we will explain that the proposed model can flexibly balance the relationship between the edge and smooth areas and achieve the goal of retaining the edge and  eliminating the staircase effect simultaneously. The  sPADMM is used to efficiently solve the proposed model.

\subsection{Model Establishment}
As mentioned above,  the harmonic model is effective in suppressing the noise in smooth areas of the image, however, it also brings about the blur to the edges inevitably. On the other hand, the total variation-based methods protect the image edges efficiently but still have a staircase artifacts in the smooth regions.
In this paper,  we combine these two models to make up for their respective shortcomings.
Note that the simple combination is not distinctive enough for image restoration.
The fixed parameter in the regularization term operates uniformly on the whole image,  while ideal edges and the smooth areas have opposite demand of regularization. Specifically, the larger parameter of regularization term is conducive to restrain noise in the smooth area,  however has a side effect of blurring the edges.
On the other hand,  a smaller regularization parameter retains edge information but brings about a staircase effect to the smooth areas.
Accordingly,  we introduce two edge adaptive parameters for TV and Tikhonov regularization terms respectively, which  treat the edges and the smooth areas flexibly and hence improve the denoising performance in a reasonable way.
On this basis,  a new edge adaptive hybrid regularization (EAHR) model is proposed.
The algorithm first detects the edges of the image to build an edge information matrix.
It then adjusts parameters for TV and Tikhonov terms according to the local information of each pixel.
Overall, the Edge Adaptive Hybrid Regularization (EAHR) model for image restoration we  consider is given as follows
\begin{equation}
{\min _{\mathbf{u} }}   \|\nabla \mathbf{u}\|_{1, \bm{\alpha}_{1}} + \frac{1}{2} \|\nabla \mathbf{u}\|_{2,  \bm{\alpha}_{2}}^{2}  + \frac{\mu}{2} \|\mathbf{A\otimes u}-\mathbf{f}\|_{2}^{2},
\label{ADTV}
\end{equation}
where $\bm{\alpha}_{1}$ and $ \bm{\alpha}_{2} \in  \mathbb{R}^{M\times N} $ are spatially varying non-negative parameter matrices,
$\|\nabla \mathbf{u}\|_{1, \bm{\alpha}_{1}}$ is given by (\ref{norm 1 alpha}) with $\bm{\alpha}_{1}$  instead of $\bm{\alpha}$,  and $ \|\nabla \mathbf{u}\|_{2,  \bm{\alpha}_{2}} $ is given by
\begin{equation}
\|\nabla \mathbf{u}\|_{2,  \bm{\alpha}_{2}} = \sqrt{\sum_{(i, j)\in \Omega}\bm{\alpha}_{2}(i, j) |\nabla \mathbf{u}(i, j)|^{2}   }.
\end{equation}

In the proposed model, the edge adaptive parameters $\alpha_1$ and $\alpha_2$ for TV and Tikhonov regularization terms play key roles in the final recovered images.
%have a crucial effect.
%For performing edge detection on an image,  we could apply edge detection operators,  for instance, Roberts operator,  Sobel operator,  Laplacian operator to detect.
%}
As explained in ~\cite{2014Ana},  the edge detection operator is a function of $|\nabla \mathbf{u}|$ for detecting the image edges.
Following it, we define the edge information matrix $\mathbf{E}$ as
\begin{equation}\label{E}
\mathbf{E}(i, j)=\frac{{1}}{{1}+|[\mathbf{G}\otimes \nabla \mathbf{u}](i, j) |^{2}}, %\quad [\mathbf{G}\otimes \nabla \mathbf{u}](i, j) =  \sum_{(s, t)\in \Omega} \mathbf{G}[(i, j)-(s, t)]\times \nabla \mathbf{u}(s, t).
\end{equation}
where $\mathbf{G}$ is Gaussian kernel, and $\mathbf{\mathbf{G}\otimes  %
\nabla \mathbf{u}}$ is  the convolution of $\mathbf{G}$ with $\nabla \mathbf{ u}$  as follows
\begin{equation*}
[\mathbf{G}\otimes \nabla \mathbf{u}](i, j) =  \sum_{(s, t)\in \Omega} \mathbf{G}[(i, j)-(s, t)]\times \nabla \mathbf{u}(s, t).
\end{equation*}
In order to improve the robustness of edge adaptive algorithm, we  binarize the edge adaptive parameters according to the  edge information matrix $\mathbf{E}$.
The spatially varying non-negative parameter matrices $\bm{\alpha}_{1}$ and $ \bm{\alpha}_{2} $ in the model (\ref{ADTV}) are defined by
\begin{equation}
\label{a1}
 \bm{\alpha}_{1}(i, j)=\cases{\alpha_{1}, & $\mathbf{E}(i, j) < \tau;$\\
\theta_{1}\alpha_{1}, & $\mathbf{E}(i, j) \ge \tau,$\\},
\end{equation}
and
\begin{equation}
\label{a2}
\bm{\alpha}_{2}(i, j)=\cases{\alpha_{2}, & $\mathbf{E}(i, j) < \tau;$\\
\theta_{2}\alpha_{2}, & $\mathbf{E}(i, j) \ge \tau,$\\}
\end{equation}
%\begin{equation}\label{a2}
%\begin{aligned}
%    \bm{\alpha}_{1}(i, j) = \left\{ \begin{array}{ll}
%\alpha_{1}, \qquad \mathbf{E}(i, j) < \tau;\\
%\theta_{1}\alpha_{1}, \qquad  \mathbf{E}(i, j) \ge \tau,
%\end{array} \right. \\
%    \bm{\alpha}_{2}(i, j) = \left\{ \begin{array}{ll}
%\alpha_{2}, & \mathbf{E}(i, j) < \tau;\\
%\theta_{2}\alpha_{2}, &  \mathbf{E}(i, j) \ge \tau,
%\end{array} \right.
%\end{aligned}
%\end{equation}
where $\alpha_{1}$ and $\alpha_{2}$ are positive parameters,  $\theta_{1}, \theta_{2} \in (0, 1)$ are  scaling parameters,  and $\tau>0$ is a threshold.
With $\bm{\alpha}_{1}$ and $ \bm{\alpha}_{2} \in  \mathbb{R}^{M\times N}$ fixed,  the proposed model is convex. Therefore, it can be solved efficiently by sPADMM.

The proposed model (\ref{ADTV}) balances the relationship between the edge and smooth areas well by adopting the two parameters  based on the edge information. 
By the proposed weights, it achieves the goal of retaining the edge and  eliminating the staircase effect simultaneously. Moreover,
while the TV-based models sometimes erroneously treats the noise in the smooth area as the image edge,   the proposed model can effectively eliminate these misjudged pixels by edge detection process.

\subsection{Algorithm}%\ref{alg5}

As mentioned, our proposed model can be efficiently solved by sPADMM. To apply this algorithm,  we first introduce an auxiliary variable $\mathbf{k}$ to take the place of  $\nabla \mathbf{u}$ in the both terms of $\|\nabla \mathbf{u}\|_{1, \bm{\alpha}_{1}}$ and $\|\nabla \mathbf{u}\|_{2,  \bm{\alpha}_{2}}^{2}$,  then the model (\ref{ADTV}) can be reformulated as the following constrained optimization problem
 \begin{equation}\label{yue}
{\min _\mathbf{u}}J(\mathbf{u})={\min_{\mathbf{u}}}  \|\mathbf{k}\|_{1, \bm{\alpha}_{1}} + \frac{1}{2} \|\mathbf{k}\|_{2,  \bm{\alpha}_{2}}^{2}  + \frac{\mu}{2} \|\mathbf{A\otimes u}-\mathbf{f}\|_{2}^{2},\quad s.t. \quad \mathbf{ k}=\nabla \mathbf{u}.
\end{equation}
The augmented Lagrangian function of the problem~(\ref{yue})~ is defined as
\begin{equation}
\fl L(\mathbf{u}, \mathbf{k}; \bm{\lambda})= \|\mathbf{k}\|_{1, \bm{\alpha}_{1}} + \frac{1}{2} \|\mathbf{k}\|_{2,  \bm{\alpha}_{2}}^{2}  + \frac{\mu}{2} \|\mathbf{A\otimes u}-\mathbf{f}\|_{2}^{2}
+{\langle\bm{\lambda}, \mathbf{k}-\nabla \mathbf{u}\rangle}+\frac{\beta}{2}\|\mathbf{k}-\nabla\mathbf{ u}\|^{2},
\label{Model k}
\end{equation}
where $\bm{\lambda}=[\bm{\lambda}_{1}, \bm{\lambda}_{2}] \in \mathbb{R}^{2M\times N} $ is the Lagrange multiplier parameter matrix with $ \bm{\lambda}(i, j) = (\bm{\lambda}_{1}(i, j),  \bm{\lambda}_{2}(i, j))  $, and $\beta>0$ is the penalty parameter.
In each iteration of   sPADMM,  we minimize $L$ with respect to $\mathbf{u}$ for fixed $\mathbf{k}$ and then minimize $L$  with respect to $\mathbf{k}$ for fixed $\mathbf{u}$.  After these two steps,  we update the multiplier  $\lambda$.
Hence, the solution to the problem (\ref{Model k}) is approached by iterating the following three  equations:
\begin{equation}\label{u}
\mathbf{u}^{n}=\mathop {\arg \min }\limits_{\mathbf{u}} L(\mathbf{u}, \mathbf{k}^{n-1};\bm{\lambda}^{n-1})+\frac{1}{2}\|\mathbf{u}-\mathbf{u}^{n-1}\|_{\mathbf{S}_{1}}^{2},
\end{equation}
\begin{equation}\label{k}
\mathbf{k}^{n}=\mathop {\arg \min }\limits_{\mathbf{k}} L(\mathbf{u}^{n}, \mathbf{k};\bm{\lambda}^{n-1})+\frac{1}{2}\|\mathbf{k}-\mathbf{k}^{n-1}\|_{\mathbf{S}_{2}}^{2},
\end{equation}
\begin{equation}
\bm{\lambda}^{n}=\bm{\lambda}^{n-1}+\eta \beta(\mathbf{k}^{n}-\nabla \mathbf{u}^{n}).
\label{solution lambda}
\end{equation}
Here,  for any $\mathbf{x}\in \mathbb{R}^{M\times N} $, $\mathbf{S} \in \mathbb{R}^{M\times N} $ is the self-adjoint positive semidefinte matrix and $\mathbf{S}\mathbf{x}$ is the matrix multiplication of $\mathbf{S}$ and $\mathbf{x}$, $\|\mathbf{x}\|_{\mathbf{S}}=\sqrt{\langle \mathbf{x},\mathbf{S}\mathbf{x}\rangle}$ denotes the matrix norm.
If we take  $\mathbf{S}_{1}=0$ and  $\mathbf{S}_{2}=0$,   the sPADMM will be the alternating direction method of multipliers (ADMM)~\cite{2010Distributed}.  In our algorithm,   $\mathbf{S}_{1}$ and $\mathbf{S}_{2}$ are positive definite   matrices. 
The sPADMM for our EAHR model is given by Algorithm \ref{alg5}.

\begin{algorithm}[H]
\label{alg5}
 \caption{sPADMM for the EAHR model}
 \KwIn{Noisy \& Blurry image $\mathbf{f}$, standard deviation $\sigma$, blurring function $\mathbf{A}$, Gaussian kernel $\mathbf{G}$}
 %Intialize: 1)$\mathbf{S}_{1}, \mathbf{S}_{2}, \alpha_{1}, \alpha_{2}, \theta_{1}, \theta_{2}, \tau, \mu, \beta, \epsilon, MaxIter$\;
 	\quad Intialize: {1)  $\mathbf{S}_{1}, \mathbf{S}_{2}, \alpha_{1},  \alpha_{2},  \theta_{1},  \theta_{2},  \tau,  \mu,  \beta,  tol$, $MaxIter$\\
			\quad \quad \quad \quad \quad 2) BM3D \cite{dabov2007image,dabov2007color} preprocessed blurred image $\mathbf{f}$\\
			\quad \quad \quad \quad \quad 3)$\mathbf{u}^{0}=0;\mathbf{k}^{0}=0;\bm{\lambda}^{0}=0$\\
}
%  \eIf{understand}
%  {
%     go to next section\;
%     current section becomes this one\;
%  }
%  {
 %  go back to the beginning of current section\;
%  }
% }
	\For {~$n = 1: MaxIter$}
{
Update
 %\begin{eqnarray*}
  %\begin{cases}
  $\cases{ 
  \mathbf{u}^{n}=\mathop {\arg \min }\limits_{\mathbf{u}} L(\mathbf{u}, \mathbf{k}^{n-1};\bm{\lambda}^{n-1})+\frac{1}{2}\|\mathbf{u}-\mathbf{u}^{n-1}\|_{\mathbf{S}_{1}}^{2},\\
  %\mathbf{u}^{n}   \mathrm{\, \,  by \, \,     \mathbf{u}=\mathcal{F}^{-1} \left\{ \frac{\mu %\mathcal{F}(\mathbf{A}^{T}).*\mathcal{F}(\mathbf{f})+\mathcal{F}(\nabla^{T}(\bm{\lambda}^{n-1}+\beta
%	\mathbf{k}^{n-1}))+\mathcal{F}(\mathbf{S}_{1}\mathbf{u}^{n-1})}{\mu \mathcal{F}(A^{T}).* \mathcal{F}(A)+\mathcal{F}(\beta \Delta +\mathbf{S}_{1})}\right\} }
%	\\
\mathbf{k}^{n}=\mathop {\arg \min }\limits_{\mathbf{k}} L(\mathbf{u}^{n}, \mathbf{k};\bm{\lambda}^{n-1})+\frac{1}{2}\|\mathbf{k}-\mathbf{k}^{n-1}\|_{\mathbf{S}_{2}}^{2},\\
   %	\mathbf{k}^{n}\mathrm{\, \,  by\, \,   \mathbf{k^{*}}=\frac{|\mathbf{k}^{*}|}{|\mathbf{n}|}\mathbf{n}}\\
   \bm{\lambda}^{n} \mathrm{\, \,  by\, \,   \bm{\lambda} = \bm{\lambda}^{n-1}+\eta \beta(\mathbf{k}^{n}-\nabla \mathbf{u}^{n})}\\
   \bm{\alpha}_{1}\, \,  \mathrm{and}  \, \, \bm{\alpha}_{2}  \mathrm{\, \,  by\, \,
   \bm{\alpha}_{m}(i, j)=\cases{\alpha_{m}, & $\mathbf{E}(i, j) < \tau;$\\
\theta_{m}\alpha_{m}, & $\mathbf{E}(i, j) \ge \tau$\\} where \,\,m=1, 2}\\
  }$%\end{cases}
 %\end{eqnarray*}

\If{$
\min \left\{\frac{\|{\mathbf{u}}^{n}-{\mathbf{u}}^{n-1}\|_{2}}{\|\mathbf{f}\|_{2}}, \frac{\| \mathbf{k}^{n}-\nabla \mathbf{u}^{n}\|_{2}}{\| \nabla \mathbf{f} \| } \right\}\leq tol,
$}
  {
     break
  }
 }
 \KwOut{Restored image~$ \mathbf{u}^{n}$}
\end{algorithm}

In each iteration of sPADMM,  the subproblems (\ref{u}) and (\ref{k}) need to be solved.
The $\mathbf{u}-$subproblem is written as
\begin{equation}
\label{u1}
\eqalign{	\min\limits_{\mathbf{u}}
	&\bigg\{\frac{\mu}{2} \|\mathbf{A\otimes u}-\mathbf{f}\|_{2}^{2}+\langle\mathbf{\lambda}^{n-1}, \mathbf{k}^{n-1}-\nabla \mathbf{u}\rangle \cr
&\quad +\frac{\beta}{2}\|\mathbf{k}^{n-1}-\nabla \mathbf{u}\|^{2}	
  +\frac{1}{2}\|\mathbf{u}-\mathbf{u}^{n-1}\|_{\mathbf{S}_{1}}^{2}\bigg\}.}
\end{equation}
%\textcolor{red}{
%For convenience of calculations, 
In our experiment, 
we choose the
self-adjoint positive definite linear operators $\mathbf{S}_{1}= \rho_{1} \mathbf {I}$, where $ \mathbf{I} $ is the unit matrix and $\rho_1>0$.
%}
Note that the subproblem (\ref{u1}) is a quadratic optimization problem subject to the optimality condition
\begin{equation*}
%\begin{aligned}
\mu \mathbf{A}^{T}(\mathbf{A\otimes u}-\mathbf{f})-\nabla^{T}\bm{\lambda}^{n-1}-\beta \nabla^{T}\mathbf{k}^{n-1}
+\beta \Delta  \mathbf{u}+\rho_{1}(\mathbf{u}-\mathbf{u}^{n-1})=0,
%\end{aligned}
\end{equation*}
which is solved by Fourier transform ~\cite{2008Anew} as follows
\begin{equation}
\mathbf{u}=\mathcal{F}^{-1} \left\{ \frac{\mu \mathcal{F}(\mathbf{A}^{T}).*\mathcal{F}(\mathbf{f})+\mathcal{F}(\nabla^{T}(\bm{\lambda}^{n-1}+\beta
	\mathbf{k}^{n-1}))+\mathcal{F}(\rho_{1}\mathbf{u}^{n-1})}{\mu \mathcal{F}(A^{T}).* \mathcal{F}(A)+\mathcal{F}(\beta \Delta +\rho_{1}\mathbf{I})}\right\}.
\label{solution u}
\end{equation}
Next,  we analyze the nonsmooth subproblem (\ref{k}) in detail to find its global minimizer.

\subsection{Solving subproblem %(17)
	(\ref{k})
}

In order to solve  subproblem (\ref{k}),  we first introduce a proposition as follows.

{\bf Proposition 2.1}.
\emph{\quad For
	any ~$\alpha, \beta >0$,  and $z, t \in \mathbb{R}^{2}$~,  the minimizer of
	\begin{equation}\label{wt}
	{\min _{z\in \mathbb{R}^{2}}} \left\{ f(z)=\alpha |z| + \frac{\beta}{2} |z-t|^{2}\right\}
	\end{equation}
	is given by
	\begin{equation}
	z^{*}%=\arg \min\limits_{s} f(s)
	=\left(1-\frac{\alpha}{\beta |t|}\right)_{+}t,
	\label{Eq: solution s}
	\end{equation}
		where $\left(a\right)_{+}= \max \left(a,0\right)$, 
 	$\frac{\mathbf{0}}{0}=\mathbf{0}$ and $ \mathbf{0}=(0, 0)$.}

Proof: \textit{
	If $t \neq \mathbf{0}$,  we have
	\begin{equation*}
	\frac{\partial f}{\partial z} =\alpha \frac{z}{|z|}+\beta (z-t).
	\end{equation*}
	Let ~$\frac{\partial f}{\partial z} = 0$,
	we get
	\begin{equation}%\label{1}
	\label{eq s norm}
	t=z+\frac{\alpha}{\beta} \frac{z}{|z|}.
	\end{equation}
	The equation (\ref{eq s norm}) implies that 
	\begin{equation}
	%|s|=|t|-\frac{\alpha}{\beta}
	%\geq 0.
	|z|=\cases{|t|-\frac{\alpha}{\beta}, & $|t|-\frac{\alpha}{\beta}\geq 0$;\\
0, & $|t|-\frac{\alpha}{\beta}<0$.\\}
	\label{2}
	\end{equation}
	It follows from (\ref{2})~ and ~(\ref{eq s norm}) that
	%\begin{equation}\label{3}
	\begin{equation*}
	z=\left(1-\frac{\alpha}{\beta |t|}\right)_{+} t.
	\end{equation*}
	%\end{equation}
	Then the  equation (\ref{Eq: solution s}) holds when $t \neq \mathbf{0}$.
	The equation (\ref{Eq: solution s}) also holds when $t = \mathbf{0}$. Therefore,  Proposition 4.1 is proved.}

The subproblem (\ref{k}) is written as
\begin{equation}
\eqalign{{\min_{\mathbf{k}}}\bigg\{\|\mathbf{k}\|_{1, \bm{\alpha}_{1}}+\frac{1}{2}\|\mathbf{k}\|_{2, \bm{\alpha}_{2}}^{2}+\frac{\mu}{2} \|\mathbf{A\otimes u}^{n}-\mathbf{f}\|_{2}^{2}
+\langle\mathbf{\lambda}^{n-1}, \mathbf{k}-\nabla\mathbf{u}^{n}\rangle \cr
\quad\quad\quad  +\frac{\beta}{2}\|\mathbf{k}-\nabla\mathbf{u}^{n}\|^{2}+\frac{1}{2}\|\mathbf{k}-\mathbf{k}^{n-1}\|_{\mathbf{S}_{2}}^{2}\bigg\}.}
\end{equation}
The  problem above is equivalently expressed as
\begin{equation*}
\eqalign{{\min _{\mathbf{k}(i, j)}} \bigg\{\bm{\alpha}_{1}(i, j) |\mathbf{k}(i, j)|+ \bm{\alpha}_{2}(i, j)|\mathbf{k}(i, j)|^{2}
+{\langle\bm{\lambda}^{n-1}(i, j), [\mathbf{k}-\nabla \mathbf{u}^{n}](i, j)\rangle} \cr
 \quad\quad\quad +  \frac{\beta}{2}\left([\mathbf{k}-\nabla \mathbf{u}^{n}](i, j)\right)^{2}
+\frac{1}{2}\|\mathbf{k}(i, j)-{\mathbf{k}(i, j)}^{n-1}\|_{\mathbf{S}_{2}}^{2}\bigg\}.}
\end{equation*}
In our experiment, %For convenience, %\textcolor{red}{ 
we use the
self-adjoint positive definite linear operators  $\mathbf{S}_{2}= \rho_{2} \mathbf {I}$, where $ \mathbf{I} $ is the unit matrix %} 
and $\rho_2>0$. Letting
$z=\mathbf{k}(i, j)$,
$q=\nabla \mathbf{u}(i, j)$,
$p_{1}=\bm{\alpha}_{1}(i, j)$,
$p_{2}=\bm{\alpha}_{2}(i, j)$
and $\widetilde{\lambda}=\bm{\lambda}(i, j)$, we rewrite the problem above as follows
\begin{equation}
\min_{z}\bigg\{p_{1}|z| + p_{2}|z|^2 + {\langle\widetilde{\lambda},  z-q \rangle} + \frac{\beta}{2}|z-q|^2+\frac{\rho_{2}}{2}\|z-z^{n-1}\|^{2}\bigg\}.
\label{Eq: min_z}
\end{equation}
where ~$|z|=\sqrt{z_{1}^{2}+z_{2}^{2}}$.
Let 
\begin{equation}
t=\frac{\beta}{\beta+ 2 p_{2}+\rho_{2}}\left(q-\frac{\tilde{\lambda}}{\beta}+\frac{\rho_{2}z^{n-1}}{\beta}\right),
\end{equation}
the problem (\ref{Eq: min_z}) is equivalent to
\begin{equation}
{\min _z} \left\{ p_{1}|z|+\frac{\beta+ 2 p_{2}+\rho_{2}}{2}|z-t|^{2} \right\}.
\end{equation}
From Proposition ~2.1,  the optimal solution of this problem is
\begin{equation}\label{kij}
z^{*}=\left( 1-\frac{p_{1}}{ (\beta + 2p_{2} + \rho_{2})|t| } \right)_{+} t.
\end{equation}
By the definition of $z^{*}$ and $t$, we have 
\begin{equation} \label{solution k}
    \mathbf{k}^{n}(i,j) = \left( 1-\frac{\bm{\alpha}_{1}(i,j)}{ (\beta + 2\bm{\alpha}_{2}(i,j) + s_{2})|t(i,j)| } \right)_{+} t(i,j),
\end{equation}
where
\begin{equation}
    t(i,j) = \frac{\beta}{\beta+ 2\bm{\alpha}_{2}(i,j)+\rho_{2}}\left(\mathbf{\nabla u(i,j)}-\frac{\mathbf{\lambda}}{\beta}+\frac{\rho_{2}\mathbf{k}^{n-1}(i,j)}{\beta}\right).
\end{equation}
Then $\mathbf{k}^{n}$ with $\mathbf{k}^{n}(i,j)$ given by (\ref{solution k}) is the solution of (\ref{k}).

%\textcolor{red}{From the above Proposition, the solution of the subproblem (\ref{k}) is
%\begin{equation}
%  \mathbf{k^{*}}=\frac{|\mathbf{k}^{*}|}{|\mathbf{n}|}\mathbf{n},
%\end{equation}
%where $$\mathbf{n}=\frac{\beta}{\beta+ 2\mathbf{\alpha}_{2}+\mathbf{S}_{2}}(\mathbf{\nabla %u}-\frac{\mathbf{\lambda}}{\beta}+\frac{\mathbf{S}_{2}\mathbf{k}^{n-1}}{\beta}),$$
%and
%$$ |\mathbf{k^{*}}| = max(n-\frac{\mathbf{\alpha_{2}}}{\beta + %2\mathbf{\alpha_{2}}+\mathbf{S}_{2}}, 0).$$
%}
%\begin{theorem}%\label{dingli1}
%Let $\{(\mathbf{u}^{n}, \mathbf{k}^{n}, \mathbf{\lambda}^{n})\}$ be the sequence generated by the algorithm sPADMM. If $\eta\in(0, \frac{1+\sqrt{5}}{2})$,  then the sequence $\{(\mathbf{u}^{n}, \mathbf{k}^{n}, \mathbf{\lambda}^{n})\}$ converges to the KKT point of (\ref{yue}).
%\end{theorem}

%\begin{CJK*}{GBK}{song}
%\textcolor{red}{ To make the paper self-contained, we establish the linear-rate convergence of algorithm 1 for the minimization problem (\ref{yue}). ÏêÏ¸%Ö¤Ã÷¹ý³ÌÔÚ¸½Â¼¡£(The detailed certification process is in the appendix.)}
%\end{CJK*}

\subsection{The linear-rate convergence}

The pseudo-code of the proposed algorithm is given in Algorithm 1. The linear-rate convergence of the algorithm is characterized in Theorem 2.1. %Please refer to Appendix for the proof.

{\bf Theorem 2.1}.
\emph{\quad Let $\{(\mathbf{u}^{n}, \mathbf{k}^{n}, \mathbf{\lambda}^{n})\}$ be the sequence generated by the algorithm sPADMM. If $\eta\in(0, \frac{1+\sqrt{5}}{2})$,  then the sequence $\{(\mathbf{u}^{n}, \mathbf{k}^{n}, \mathbf{\lambda}^{n})\}$ converges to the KKT point of (\ref{yue}).}

%{\bf Note:} 
The detailed proof of the theorem can be found in the Appendix.

\section{Numerical Simulation}

\begin{figure}[t]
\centering
 \begin{tabular}{c@{\hskip 2pt}c@{\hskip 2pt}c@{\hskip 2pt}c@{\hskip 2pt}c@{\hskip 2pt}c@{\hskip 2pt}c@{\hskip 2pt}c@{\hskip 2pt}c@{\hskip 2pt}c@{\hskip 2pt}c@{\hskip 2pt}c@{\hskip 2pt}c}
\includegraphics[width=0.23\textwidth]{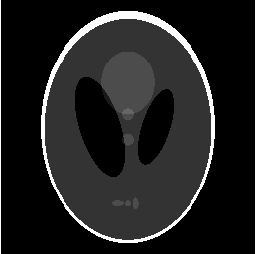}&
\includegraphics[width=0.23\textwidth]{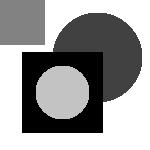}&
\includegraphics[width=0.23\textwidth]{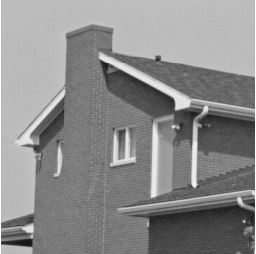}&
\includegraphics[width=0.23\textwidth]{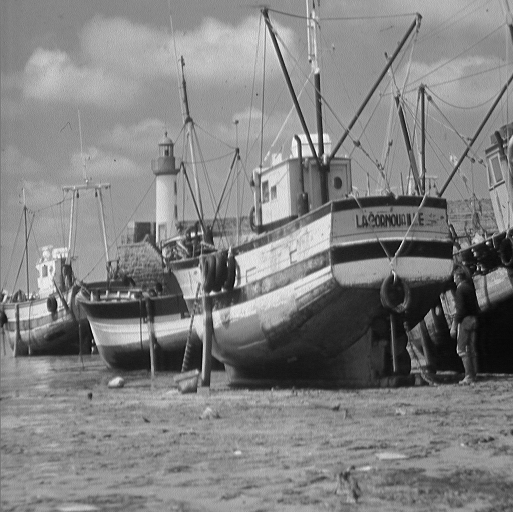}&\\
\scriptsize Shepp-Logan & \scriptsize Shape150&\scriptsize House&\scriptsize Boat&\\
\includegraphics[width=0.23\textwidth]{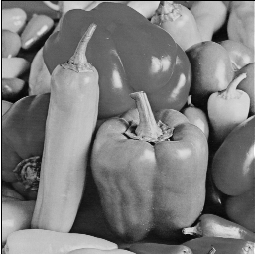}&
\includegraphics[width=0.23\textwidth]{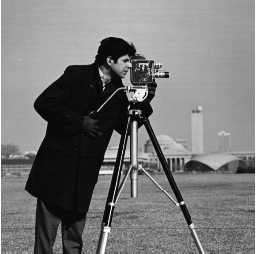}&
\includegraphics[width=0.23\textwidth]{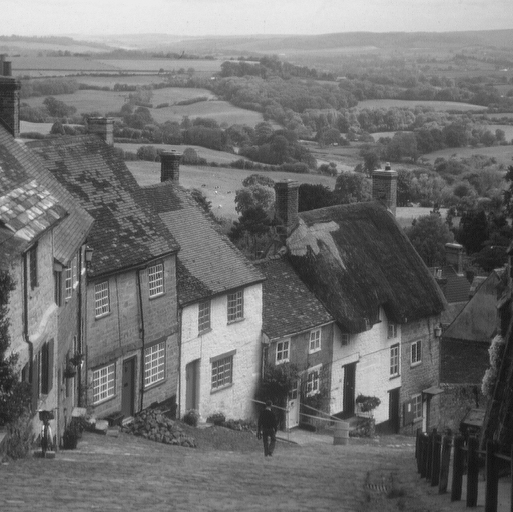}&
\includegraphics[width=0.23\textwidth]{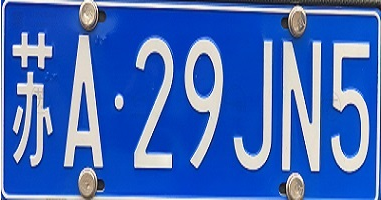}&\\
\scriptsize Pepper&\scriptsize Cameraman&\scriptsize Hill&\scriptsize Plate&\\
\includegraphics[width=0.23\textwidth]{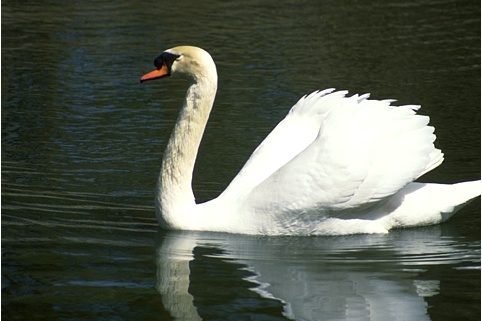}&
\includegraphics[width=0.23\textwidth]{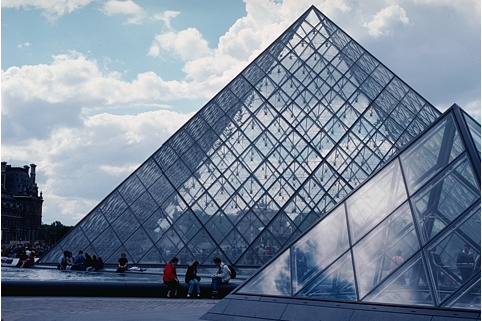}&
\includegraphics[width=0.23\textwidth]{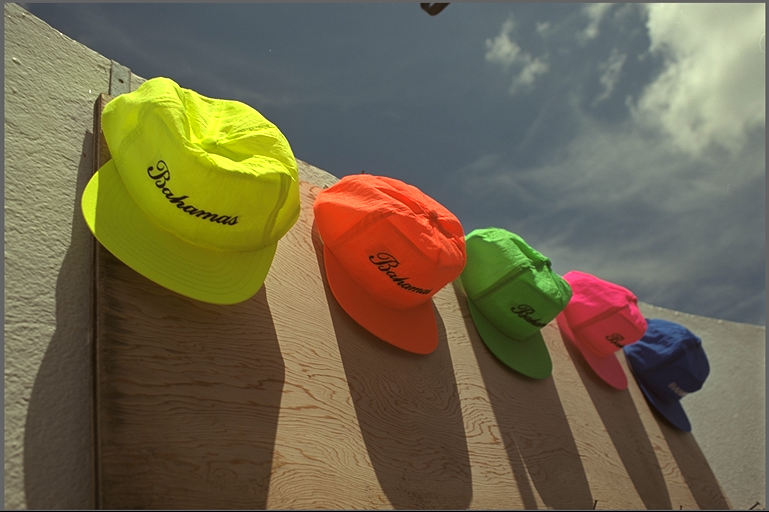}&
\includegraphics[width=0.23\textwidth]{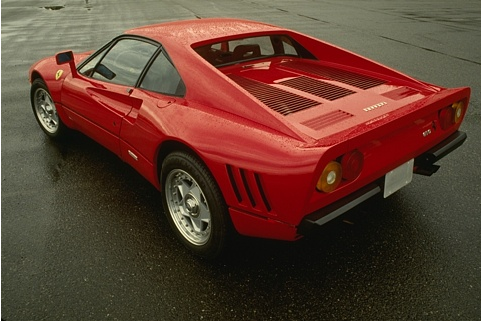}\\
\scriptsize Duck&\scriptsize Building&\scriptsize Hats&\scriptsize Car
\end{tabular}
\caption{\small{ Set of 12 testing images. %Shepp-Logan and Shape150 are simulated images. 
House, Boat, Pepper, Cameraman, Hill  from the Tampere University of Technology(\url{http://www.cs.tut.fi/~foi/GCF-BD3D/}). Plate from [22].  Duck, Building, Car from the Berkeley segmentation database. Hats from Kodak Image Dataset(\url{http://www.cs.albany.edu/~xypan/research/snr/Kodak.html}).}}\label{fig:test}
\end{figure}

In this section,  we present simulation results with twelve testing images,  including 7 gray images and 5 color images,  as  shown in Figure \ref{fig:test}.
We consider three types of blurring functions  Gaussian blur (GB),  Motion blur (MB) and Average blur (AB) with different levels of additive Gaussian noise.
%\textcolor{red}{For the case of Gaussian noise, the deviation $\sigma$ = 3, 5, 8 and 10.
%For Gaussian blur, we use GB(9,5). MB(20, 60) is represents Motion blur, where 20 is the length and 60 is the angle.
% AB(9,9) is considered that the size of Average blur kernel is $9\times9$.}
%In addition,  Gaussian noise with standard deviation $\sigma$ is added to the blurred images.
For instance, the notation $\mathrm{GB(9, 5)}/\sigma=5$ represents Gaussian kernel %$ \exp\left( \frac{\|(i,j)-(s,t)\|_{2}^{2}}{2\times 5^{2}}\right)  $
with the  free parameter $5$  and size $9\times9$, and additive Gaussian noise with standard deviation $\sigma=5$. %
%Similarly, $\mathrm{MB(20, 60)}/\sigma=5$ is denoted the motion blur of size $20\times20$ with additive Gaussian noise of standard deviation $\sigma=5$.
%We use $\mathrm{AB(9, 9)}/\sigma=3$ for average blur of size $9\times9$ and Gaussian noise of the standard deviation $\sigma=3$.

To illustrate the effectiveness of the proposed model,  we compare our model with the classic TV ~\cite{1992Nonlinear},  DCA with $L_{1}-0.5L_{2}$ ~\cite{2015Aw},  TRL2 ~\cite{2018Ag}, SOCF ~\cite{2020Ani} and BM3D~\cite{2008Image}.
All the experiments are performed under Windows 10 and MATLAB R2018a running on a desktop (Intel(R) Core(TM) i5-8250 CPU @1.60 GHz).
The termination criterion for all experiments is defined as follows
\begin{equation*}
\min \left\{\frac{\|{\mathbf{u}}^{n}-{\mathbf{u}}^{n-1}\|_{2}}{\|\mathbf{f}\|_{2}}, \frac{\| \mathbf{k}^{n}-\nabla \mathbf{u}^{n}\|_{2}}{\| \nabla \mathbf{f} \| } \right\}\leq tol,
\end{equation*}
where $n$ is the number of iterations,  the tolerance value $tol$ is set as $5e-5$.
Quantitatively,  we evaluate the quality of image restoration by the peak signal to noise ratio (PSNR) and structural similarity index (SSIM). The SSIM is defined by Wang et al. ~\cite{2004Image} and PSNR is calculated by
\begin{equation*}
\mathrm{PSNR} = 10\log_{10}\frac{255^2}{\mathrm{MSE}}, \quad \mathrm{MSE} = \frac{1}{M\times N} \sum_{(i, j)\in\Omega}(\mathbf{u}(i, j)-\hat{\mathbf{u}}(i, j))^{2},
\end{equation*}
where
% \begin{equation*}
% \mathrm{MSE} = \frac{1}{M\times N} \sum_{(i, j)\in\Omega}(\mathbf{u}(i, j)-\hat{\mathbf{u}}(i, j))^{2},
% \end{equation*}
$M \times N$ is the image size,  $\mathbf{u}$ is the original image and $\hat{\mathbf{u}}$ is the restored image.

\subsection{Parameter setting}
In this section, we introduce all the parameters used in our algorithm.
From (\ref{a2}),  $\theta_{1}, \theta_{2} \in (0, 1)$ are  scaling parameters,  and $\tau>0$ is a threshold  which is used to adjust the value of $\alpha_{1}$ and $\alpha_{2}$.
 There are three parameters $(\mu,  \beta,  \tau)$ in our Algorithm \ref{alg5} that  need to be adjusted.
 We choose the self-adjoint positive definite linear operators $\mathbf{S}_{1}=\mathbf{S}_{2}=\rho \mathbf{I}$ with $\rho = 0.1$ and the maximum iterative number (denoted by $MaxIter$ ) is set as $500$. Through a multitude of experiments,  we choose better parameters to improve the quality of restoring images damaged by blur and noise.

%\textcolor{red}{
We have done a lot of simulations to find the best value of parameters $\mu$, $\tau$ and $\beta$ for each blurring kernel with different noise levels.
As a result, we get the parameters $\mu$, $\tau$ and $\beta$ as functions of $\sigma \in (0,100]$ for each blurring kernel. 
%Since both the blurring kernel and Gaussian noise level used in this paper are different, the choices of corresponding parameters are different. We use parameter values and noise values for polynomial fitting, and get their functional relationship. 
Taking Gaussian blur as a example, 
the parameters  $\mu$, $\tau$ and $\beta$ are given as: 
\begin{equation*}%\label{19}
%\eqalign{ &\mu = 170{\sigma}^{2}-2690\sigma+11470,\\
\eqalign{ 
\mu = 1.8\sigma^{2}\times 10^{2}-2.7\sigma\times 10^{3}+1.1\times 10^{4},\cr
%&\tau=0.003448\sigma+0.9026,\\
\tau=3.4 \sigma\times 10^{-3}+0.9,\cr
%&\beta=−0.0003711\sigma+0.1175,  %while \sigma\leq 100.
\beta =-3.7\sigma\times 10^{-4}+1.2\times 10^{-1},
}
\end{equation*}
where $\sigma \in (0,100]$.
%$\mu = 170{\sigma}^{2}-2690\sigma+11470$,
%$\tau=0.003448\sigma+0.9026$, 
%$\beta=-0.0075{\sigma}^{2}+0.1035\sigma-0.1392$. 
When  $\sigma=3$, the corresponding values of $\mu$, $\tau$ and $\beta$ are 5000, 0.9 and 0.1 respectively. In the case of Motion blur and Average blur, 
%the functional relationship between noise and parameters is similar to that of Gaussian blur, 
we have similar functions for the parameters $\mu$, $\tau$ and $\beta$
which are given in the code.
%}

%\textcolor{red}{
The parameter settings for the other compared algorithms are given below. The
TV model~\cite{1992Nonlinear} has two parameters $\mu$ and $\lambda$, where $\mu$ is the parameter to balance
the data fidelity and regularity terms and  $\lambda$ plays an important role in  analyzing the convergence.
In our tests, we let $\mu\in\{200, 300, 400, 700, 1500\}$ and $\lambda\in\{10, 15, 20, 50, 80, 100, 200, 300\}$.
For the DCA model~\cite{2015Aw}, we fix  the parameter $\lambda=1$, and then choose the parameter $\mu$ from the
sequence $\{50, 100, 150,200, 300, 380, 700, 800, 1500\}$. The parameters used by the TRL2 model~\cite{2018Ag} are set
as $\alpha\in \{5, 10, 15, 20, 30, 50, 100, 150, 200\}$, $\tau\in\{0.1, 0.2, 0.3, 0.5\}$ and $\beta\in\{0.1, 0.5, 1, 2, 3, 5, 10\}$.
In the SOCF model~\cite{2020Ani}, we take fixed values  $\mu_{1}=0.2, \mu_{2}=1e-1$ and $\eta=1.6$
%, which $\eta$ is
represented the step-length. We choose $\mu\in\{1e-4, 1e-3\}$ and $\lambda\in\{2, 3, 4, 7, 11,13\}$.  The BM3D model~\cite{2008Image} has two parameters: regularized inversion (RI) which is the  collaborative hard-thresholding for BM3D filtering and regularized Wiener inversion (RWI) for collaborative Wiener filtering. We set these two parameters as $RI\in \{4e-4, 4e-3, 1e-3\}$ and $RWI\in \{5e-3, 3e-2\}$. 
%}
 %For Gaussian blur $\mathrm{GB(9, 5)}/\sigma=5$, the images corrupted by blur and noise  restoration effect is better while choosing $\mu=2100$,  $\beta=0.25$ and $\tau=0.95$ for all images. Similarly,  we set $\mu=3500$,  $\beta=0.2$,  $\tau=0.9$  for Motion blur  $\mathrm{MB(20, 60)}/\sigma=5$ and $\mu=6000$,  $\beta=0.2$,  $\tau=0.9$ for Average blur $\mathrm{AB(20, 60)}/\sigma=3$.

\subsection{Experimental results}
In order to verify the performance of our proposed model,  we have tested the recovery results of gray images and color images under different blur kernels and different Gaussian noise levels.
Tables \ref{tab:table e}, ~\ref{tab:table g} and ~\ref{tab:table AB} show the values of PSNR,  SSIM of the classic TV ~\cite{1992Nonlinear},  DCA with $L_{1}-0.5L_{2}$ ~\cite{2015Aw},  TRL2 ~\cite{2018Ag},  SOCF ~\cite{2020Ani} and BM3D ~\cite{2008Image} on Gaussian blur $\mathrm{GB(9, 5)}/\sigma=5$ and Motion blur $\mathrm{MB(20, 60)}/\sigma=5$ and Average blur $\mathrm{AB(20, 60)}/\sigma=3$.
The number in bold means the best result.
%\textcolor{red}{
Obviously, our algorithm gives the best results in most of the cases. We also show the average of PSNR/SSIM value of the restored image  in the Table \ref{tab:table aa}.
Comprehensive evaluations on the image set with three kinds of kernels and four different noise levels demonstrate the superiority of our proposed method when compared with other state-of-the-art methods. 
%}

\begin{table}[t]\scriptsize%\footnotesize%\small
	\renewcommand\arraystretch{1.75}
\caption{\label{tab:table e}The value of PSNR and SSIM of the test images recovered by different models for  $\mathrm{GB(9, 5)}/\sigma=5$.}
%\footnotesize\rm
%\begin{indented}
\lineup
\centering
%\item[]
\begin{tabular}{@{}*{8}{c}}
%\item[]\begin{tabular}{@{}1111111}
\br
PSNR/SSIM&Degraded&TV~\cite{1992Nonlinear}&DCA ~\cite{2015Aw}&TRL2 ~\cite{2018Ag}&SOCF ~\cite{2020Ani}&BM3D~\cite{2008Image}&\0Ours\cr
\mr
Shepp-Logan&18.93/0.578&  22.77/0.930&  23.29/0.893&  23.68/0.873&  23.96/0.918&24.56/0.846  & \textbf{ 25.66}/\textbf{0.941}\cr
Shape150&18.46/0.541&  25.05/0.859&  25.66/0.866&  25.62/0.869&  25.96/0.910& 26.44/0.873  & \textbf{27.74}/\textbf{0.916}\cr
House &24.12/0.560&  27.90/0.784&  28.18/0.800&  28.05/0.676&  29.10/0.797& 29.90/0.783  & \textbf{30.66}/\textbf{0.826}\cr
Boat  &23.34/0.489&  25.03/0.633&  25.43/0.655&  26.12/0.640&  26.45/0.690&  26.76/0.700  &\textbf{26.92}/\textbf{0.704}\cr
Pepper&21.41/0.573&  23.46/0.735&  23.69/0.736&  24.71/0.705&  24.87/0.781& \textbf{ 26.42}/0.763 & 25.86/\textbf{0.793}\cr
Cameraman &20.87/0.499&  23.10/0.716&  23.49/0.734&  24.16/0.645&  24.24/0.745& \textbf{24.74}/0.737 &24.73/\textbf{0.758}\cr
Hill  &24.85/0.497&  26.86/0.631&  27.12/0.653&  27.35/0.644&  27.55/0.670&  28.07/0.688 & \textbf{28.08}/\textbf{0.693}\cr
Plate  &17.40/0.779&  21.64/0.910&  21.92/0.916&  21.90/0.912&  22.71/0.923& 24.57/0.942 & \textbf{24.84}/\textbf{0.947}\cr
Duck  &24.17/0.659&  27.00/0.789&  27.13/0.804&  27.25/0.742&  28.11/0.823&   28.60/0.830 &\textbf{29.13}/\textbf{0.834}\cr
Building &19.54/0.636&  20.17/0.701&  20.40/0.718&  20.80/0.711&  20.87/0.739&  \textbf{21.33}/0.760  &21.32/\textbf{0.762}\cr
Hats  &27.04/0.858&  28.95/0.937&  29.23/0.939&  29.10/0.916&  29.75/0.941&  30.10/0.944 & \textbf{30.22}/\textbf{0.946}\cr
Car   &23.92/0.755&  25.18/0.828&  25.70/0.838&  26.03/0.819&  26.22/0.841&  26.74/0.847 &\textbf{26.98}/\textbf{0.853}\cr
Average   &22.01/0.619&  24.76/0.788&  25.10/0.796&  25.39/0.738&  25.82/0.815& 26.52/0.809  & \textbf{26.89}/\textbf{0.830}\cr

\br
\end{tabular}
%\end{indented}
\centering
\end{table}
%\normalsize

\begin{table}[t]\scriptsize%\footnotesize%\small
\centering
	\renewcommand\arraystretch{1.75}
\caption{\label{tab:table g}The value of PSNR and SSIM of the test images recovered by different models for $\mathrm{MB(20, 60)}/\sigma=5$.}
%\begin{indented}
\lineup
%\item[]
\begin{tabular}{@{}*{8}{l}}
%\item[]\begin{tabular}{@{}1111111}
\br
PSNR/SSIM&Degraded&TV~\cite{1992Nonlinear}&DCA ~\cite{2015Aw}&TRL2 ~\cite{2018Ag}&SOCF ~\cite{2020Ani}&BM3D~\cite{2008Image}&\0Ours\cr
\mr
Shepp-Logan &17.46/0.546&  24.78/\textbf{0.942}&  25.40/0.919&  25.11/0.883&  25.43/0.870&    25.67/0.808&\textbf{27.87}/0.903\\%DCA:mu=50
		Shape150    &16.33/0.469&  25.78/0.901&  \textbf{30.52}/\textbf{0.926}&  24.24/0.602&  25.16/0.854& 26.62/0.865 & 27.94/0.869\\%DCA:mu=500
		House       &21.45/0.503&  25.21/0.730&  27.93/0.738&  25.85/0.653&  28.73/0.772&   29.24/0.786 &\textbf{29.55}/\textbf{0.793}\\
		Boat        &22.01/0.456&  25.54/0.658&  25.96/0.660&  25.73/0.652&  26.28/0.688&  26.63/0.698  &\textbf{26.89}/\textbf{0.703}\\
		Pepper     &20.06/0.532&  23.94/0.763&  24.16/0.732&  23.44/0.665&  24.86/\textbf{0.772}& \textbf{ 25.39}/0.768  &24.79/0.755\\
		Cameraman   &19.80/0.478&  23.92/0.736&  24.50/0.705&  24.41/0.612&  24.91/0.739&  25.25/0.745 & \textbf{25.29}/\textbf{0.746}\\
		Hill     &23.21/0.451&  26.61/0.632&  26.88/0.642&  26.72/0.639&  26.98/0.655&  27.42/0.670 & \textbf{27.50}/\textbf{0.675}\\
		Plate     &15.93/0.730&  20.99/0.902&  21.55/0.911&  21.19/0.903&  23.09/0.932&   24.11/0.941 &\textbf{24.44}/\textbf{0.953}\\
		Duck     &22.48/0.623&  26.79/0.782&  26.81/0.782&  26.61/0.683&  27.98/0.792&  28.03/\textbf{0.808}  &\textbf{28.57}/0.797\\
		Building    &18.93/0.628&  20.87/0.741&  20.97/0.749&  21.72/0.740&  22.30/0.796&  22.86/\textbf{0.816}  &\textbf{22.92}/0.806\\
		Hats     &25.44/0.840&  28.74/0.934&  28.78/\textbf{0.935}&  28.31/0.885&  29.29/0.927&  29.22/0.930 & \textbf{29.62}/0.931\\
		Car         &22.05/0.737&  24.97/0.830&  25.57/0.838&  25.56/0.803&  25.82/0.841&  26.49/0.843 & \textbf{26.89}/\textbf{0.844}\\
		Average     &20.43/0.583&  24.85/0.796&  25.75/0.795&  24.91/0.710&  25.90/0.803& 26.41/0.807 & \textbf{26.84}/\textbf{0.815}\\

\br
\end{tabular}
%\end{indented}
\centering
\end{table}

\begin{table}[t]
\scriptsize%\footnotesize%\small
\centering
	\renewcommand\arraystretch{1.75}
\caption{\label{tab:table AB}The value of PSNR and SSIM of the test images recovered by different models for  $\mathrm{AB(9, 9)}/\sigma=3$.}
%\begin{indented}
\lineup
%\item[]
\begin{tabular}{@{}*{8}{l}}
%\item[]\begin{tabular}{@{}1111111}
\br
PSNR/SSIM&Degraded&TV~\cite{1992Nonlinear}&DCA ~\cite{2015Aw}&TRL2 ~\cite{2018Ag}&SOCF ~\cite{2020Ani}&BM3D~\cite{2008Image}&\0Ours\cr
\mr
Shepp-Logan &18.65/0.708&24.39/0.947&25.82/0.926&24.94/0.853&	24.73/0.938	&25.71/0.884&\textbf{27.35}/\textbf{0.951} \\
		Shape150 &18.16/0.612	&26.25/0.900&28.48/0.933&27.97/\textbf{0.946}	&27.46/0.941&28.01/0.928&\textbf{28.97}/0.918\\
		House  & 23.96/0.618 &29.09/0.807&29.44/0.823&29.61/0.768&	30.45/0.825& 31.47/0.836 &\textbf{31.60}/\textbf{0.837}\\
		Boat     & 23.23/0.521 &	25.82/0.662	 &26.59/0.703 &	27.12/0.708	 &27.23/0.721 & 28.06/0.751 &	\textbf{28.14}/\textbf{0.751}\\
		Pepper    &  21.25/0.613 &	24.21/0.763	 &25.45/0.781 &	25.78/0.777	 &27.02/\textbf{0.823}&27.75/0.806	&\textbf{27.89}/0.820\\
		Cameraman  &20.70/0.561	 &23.87/0.743 &	24.47/0.766 &	24.78/0.642	 &24.66/0.704 &25.81/0.781	&\textbf{25.98}/\textbf{0.792}\\
		Hill    & 24.85/0.528	 &27.46/0.661 &	28.03/0.695 &	28.39/0.709	 &28.43/0.708 &\textbf{29.03}/0.731&28.96/\textbf{0.736}\\
		Plate  &17.08/0.766	&22.98/0.930&	23.86/0.941	&23.54/0.935	&25.04/0.950&26.19/0.958	&\textbf{26.80}/\textbf{0.963}  \\
		Duck   &24.12/0.709	&28.43/0.821&	28.59/0.837	&29.21/0.845	&29.34/0.850&29.91/0.863&\textbf{30.62}/\textbf{0.863} \\
		Building &19.47/0.650&	20.70/0.734	&20.87/0.750&	21.02/0.749	&21.48/0.778&\textbf{22.06}/0.794&22.00/\textbf{0.794}\\
		Hats  & 27.38/0.897&	29.83/0.945&	29.86/0.945	&30.02/0.935	&\textbf{31.63}/0.952&31.34/0.954	&31.53/\textbf{0.955}  \\
		Car & 23.93/0.785&	26.40/0.847&	26.43/0.850	&26.49/0.843	&27.16/0.859&27.41/0.861	&\textbf{27.67}/\textbf{0.863}      \\
		Average   &21.90/0.664&	25.79/0.813&	26.49/0.829	&26.57/0.768	&27.05/0.837&27.73/0.846	&\textbf{28.13}/\textbf{0.854} \\

\br
\end{tabular}
%\end{indented}
\centering
\end{table}

\begin{table}[t]\scriptsize%\footnotesize%\small%\tiny%
\centering
	\renewcommand\arraystretch{2.5}
\caption{\label{tab:table aa}The average value of PSNR and SSIM of the test images recovered by different models for  all experiments.}
%\begin{indented}
\lineup
%\item[]
\begin{tabular}{@{}*{9}{l}}
%\item[]\begin{tabular}{@{}1111111}
\br
Kernel&$\sigma$ & Degraded&TV~\cite{1992Nonlinear}&DCA ~\cite{2015Aw}&TRL2 ~\cite{2018Ag}&SOCF ~\cite{2020Ani}&BM3D~\cite{2008Image}&\0Ours\cr
\mr
$\mathrm{GB(9, 5)}$  &$\sigma=3$  &22.22/0.678 &27.30/0.838 &26.91/0.821 &26.26/0.785 &27.40/0.845 &27.46/0.836 &\textbf{27.82/0.857}\\
		             &$\sigma=5$  &22.00/0.619 &24.76/0.789 &25.10/0.796 &25.40/0.762 &25.82/0.812 &26.52/0.809 &\textbf{26.81/0.831}\\
		             &$\sigma=8$  &22.37/0.527 &25.37/0.791 &24.96/0.770 &24.03/0.727 &25.57/0.789 &25.65/0.786 &\textbf{25.97/0.811}\\
	                 &$\sigma=10$ &21.15/0.472 &25.17/0.771 &24.63/0.765 &24.18/0.721 &25.16/0.775 &25.24/0.776 &\textbf{25.74/0.796}\\
\mr
$\mathrm{MB(20, 60)}$&$\sigma=3$  &20.58/0.638 &27.75/0.819 &28.03/0.829 &26.93/0.812 &27.41/0.834 &27.70/0.834 &\textbf{28.09/0.846}\\
		             &$\sigma=5$  &20.43/0.583 &24.84/0.793 &25.75/0.795 &24.91/0.727 &25.90/0.799 &26.41/0.807 &\textbf{26.86/0.815}\\
		             &$\sigma=8$  &20.10/0.496 &25.26/0.768 &25.10/0.757 &23.51/0.737 &24.94/0.760 &25.24/0.780 &\textbf{25.75/0.783}\\
		             &$\sigma=10$ &19.82/0.445 &24.58/0.757 &24.63/0.763 &23.56/0.696 &24.46/0.755 &24.69/0.768 &\textbf{25.25/0.776}\\
\mr
$\mathrm{AB(9, 9)}$  &$\sigma=3$  &21.90/0.664 &25.79/0.813 &26.49/0.829 &26.57/0.793 &27.05/0.837 &27.73/0.846 &\textbf{28.13/0.854}\\
		             &$\sigma=5$  &21.70/0.605 &26.41/0.812 &26.07/0.811 &24.89/0.730 &26.65/0.816 &26.76/0.819 &\textbf{27.16/0.831}\\
		             &$\sigma=8$  &21.26/0.515 &25.61/0.792 &25.12/0.776 &24.37/0.737 &25.71/0.783 &25.87/0.793 &\textbf{26.24/0.809}\\
	                 &$\sigma=10$ &20.87/0.462 &25.12/0.767 &24.72/0.756 &24.12/0.726 &25.63/0.793 &25.44/0.781 &\textbf{25.84/0.798}\\

\br
\end{tabular}
%\end{indented}
\centering
\end{table}

\begin{table}\footnotesize%\scriptsize%\small
\centering
	\renewcommand\arraystretch{1.2}
\caption{\label{tab:table 2}The computation time(s) for all compared methods on  a desktop (Intel(R) Core(TM) i5-8250 CPU @1.60 GHz).}
%\begin{indented}
\lineup
%\item[]
\begin{tabular}{@{}*{8}{l}}
%\item[]\begin{tabular}{@{}1111111}
\br
Kernel&$Image size$&TV~\cite{1992Nonlinear}&DCA ~\cite{2015Aw}&TRL2 ~\cite{2018Ag}&SOCF ~\cite{2020Ani}&BM3D~\cite{2008Image}&\0Ours\cr
\mr
$\mathrm{GB(9, 5)}$&$512 \times 512$ gray  &\textbf{2.93}  &50.07  &7.05  &12.99  &5.73  &20.22\\
                   &$768 \times 512$  RGB &\textbf{12.96}  &162.96  &31.44  &60.56 &27.94  &79.28\\
\mr
$\mathrm{MB(20, 60)}$&$512 \times 512$ gray  &\textbf{2.46}  &60.61  &10.90  &17.03  &5.81  &20.52\\
                     &$768 \times 512$  RGB  &\textbf{12.89}  &137.92  &30.74 &93.01 &28.69  &100.66\\
\mr
$ \mathrm{AB(9, 9)}$&$512 \times 512$ gray  &\textbf{2.91}  &44.39  &11.82  &12.47  &5.64  &14.67\\
                    &$768 \times 512$  RGB &\textbf{12.76}  &152.73  &36.67  &62.05  &28.21  &85.09\\
\br
\end{tabular}
%\end{indented}
\centering
\end{table}

\begin{figure}[!ht]
\centering
%\renewcommand{\arraystretch}{0.2}\addtolength{\tabcolsep}{-5pt} \vskip3mm
%\backslash\footsize{1pt}{\baselineskip}\selectfont
 \begin{tabular}{c@{\hskip 2.5pt}c@{\hskip 2.5pt}c@{\hskip 2.5pt}cc}
    \footnotesize
\includegraphics[width=0.23\linewidth]{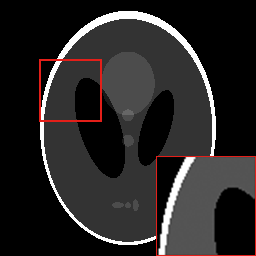}&
\includegraphics[width=0.23\linewidth]{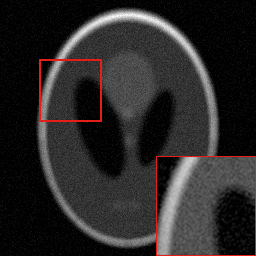}&
\includegraphics[width=0.23\linewidth]{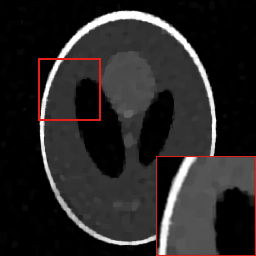}&
\includegraphics[width=0.23\linewidth]{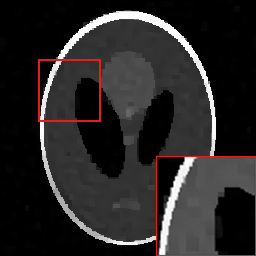}&\\
\small (a) Original image&
\small (b) Degraded image&
\small (c) TV ~\cite{1992Nonlinear}&
\small (d) DCA ~\cite{2015Aw}\\
&
\small 18.74/0.398&
\small 24.04/0.804&
\small 23.11/0.823\\
\includegraphics[width=0.23\linewidth]{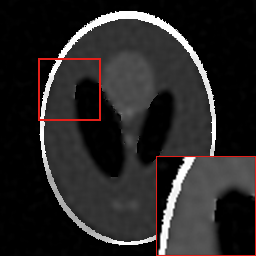}&
\includegraphics[width=0.23\linewidth]{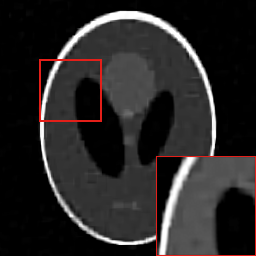}&
\includegraphics[width=0.23\linewidth]{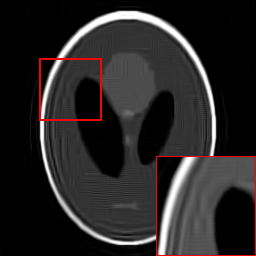}&
\includegraphics[width=0.23\linewidth]{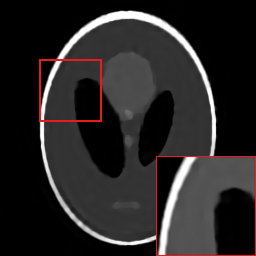}\\
\small (e) TRL2 ~\cite{2018Ag}&
\small (f) SOCF ~\cite{2020Ani}&
\small (g) BM3D ~\cite{2008Image}&
\small (h) Ours\\
\small 23.44/0.773&
\small 23.93/0.827&
\small 23.76/0.833&
\small 24.83/0.929
 \end{tabular}
%\caption{\small{ Deblurring result of GB(9, 5) and $\sigma=8$ for gray image "Shepp-Logan" with zoomed areas and PSNR values and SSIM values. }}
\caption{ Deblurring result of GB(9, 5) and $\sigma=8$ for gray image "Shepp-Logan" with zoomed areas and PSNR values and SSIM values. }
\label{fig:GBg}
\centering
\end{figure}

\begin{figure}[!ht]
\centering
%\renewcommand{\arraystretch}{0.2}\addtolength{\tabcolsep}{-5pt} \vskip3mm
%\backslash\footsize{1pt}{\baselineskip}\selectfont
 \begin{tabular}{c@{\hskip 2.5pt}c@{\hskip 2.5pt}c@{\hskip 2.5pt}cc}
    \footnotesize
\includegraphics[width=0.23\linewidth]{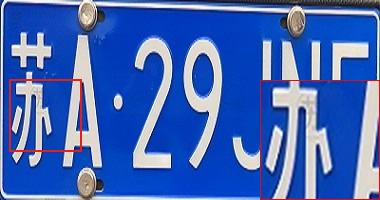}&
\includegraphics[width=0.23\linewidth]{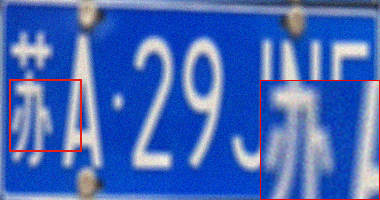}&
\includegraphics[width=0.23\linewidth]{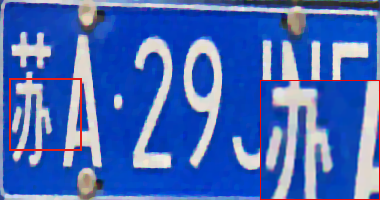}&
\includegraphics[width=0.23\linewidth]{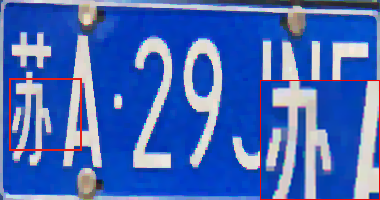}&\\
\small (a) Original image&
\small (b) Degraded image&
\small (c) TV ~\cite{1992Nonlinear}&
\small (d) DCA ~\cite{2015Aw}\\
&
\small 17.12/0.765&
\small 22.06/0.915&
\small 21.62/0.910\\
\includegraphics[width=0.23\linewidth]{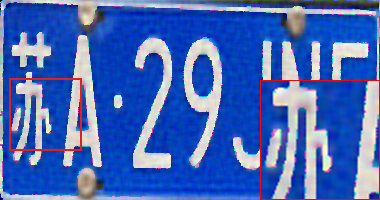}&
\includegraphics[width=0.23\linewidth]{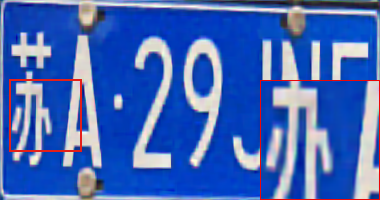}&
\includegraphics[width=0.23\linewidth]{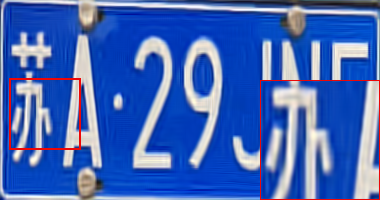}&
\includegraphics[width=0.23\linewidth]{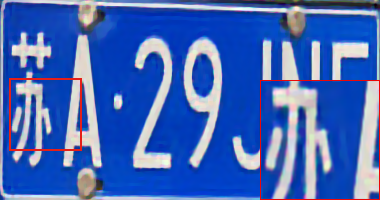}\\
\small (e) TRL2 ~\cite{2018Ag}&
\small (f) SOCF ~\cite{2020Ani}&
\small (g) BM3D ~\cite{2008Image}&
\small (h) Ours\\
\small 20.10/0.868&
\small 22.23/0.914&
\small 22.77/0.920&
\small 23.30/0.928
 \end{tabular}
\caption{ Deblurring result of GB(9, 5) and $\sigma=10$ for color image "Plate" with zoomed areas and PSNR values and SSIM values. }
\label{fig:GBc}
\centering
\end{figure}

\begin{figure}[!ht]
\centering
%\renewcommand{\arraystretch}{0.2}\addtolength{\tabcolsep}{-5pt} \vskip3mm
%\backslash\footsize{1pt}{\baselineskip}\selectfont
 \begin{tabular}{c@{\hskip 2.5pt}c@{\hskip 2.5pt}c@{\hskip 2.5pt}cc}
    \footnotesize
\includegraphics[width=0.23\linewidth]{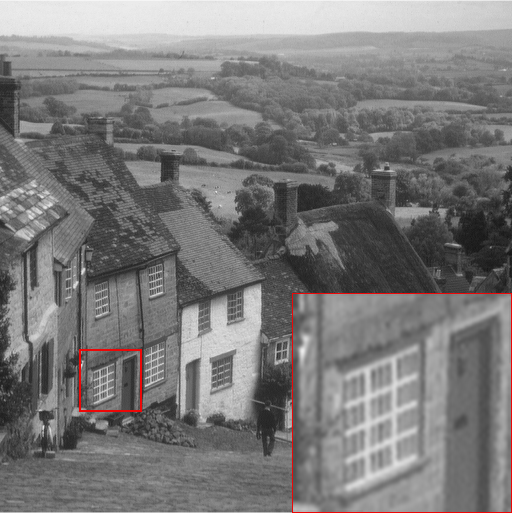}&
\includegraphics[width=0.23\linewidth]{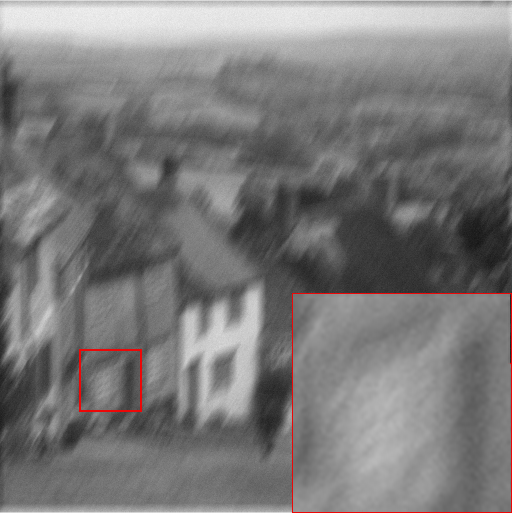}&
\includegraphics[width=0.23\linewidth]{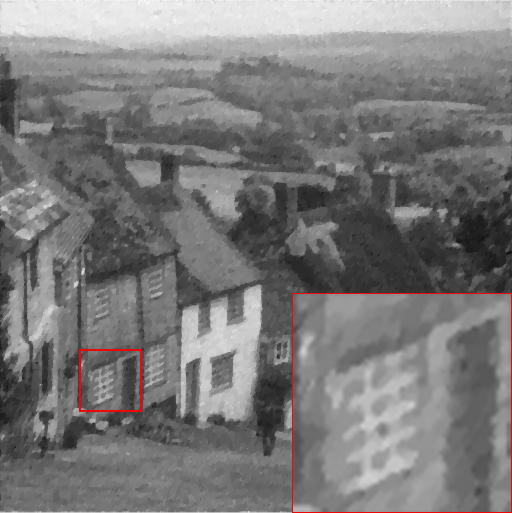}&
\includegraphics[width=0.23\linewidth]{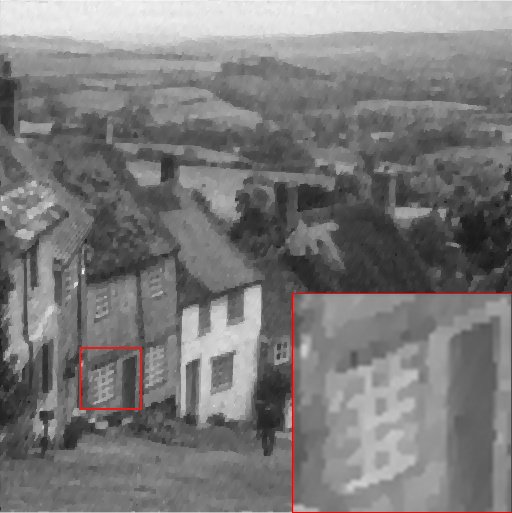}&\\
\small (a) Original image&
\small (b) Degraded image&
\small (c) TV ~\cite{1992Nonlinear}&
\small (d) DCA ~\cite{2015Aw}\\
&
\small 23.43/0.499&
\small 28.15/0.703&
\small 28.04/0.704\\
\includegraphics[width=0.23\linewidth]{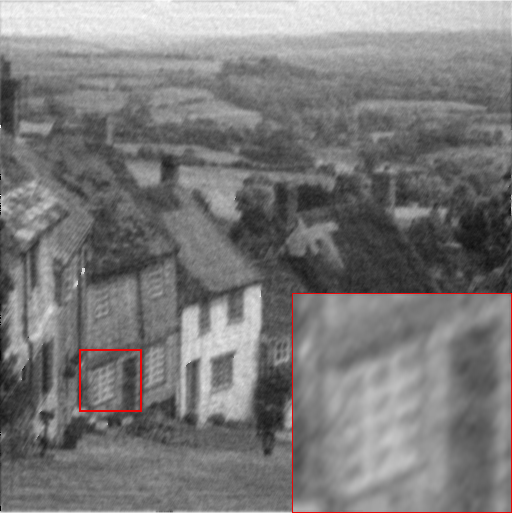}&
\includegraphics[width=0.23\linewidth]{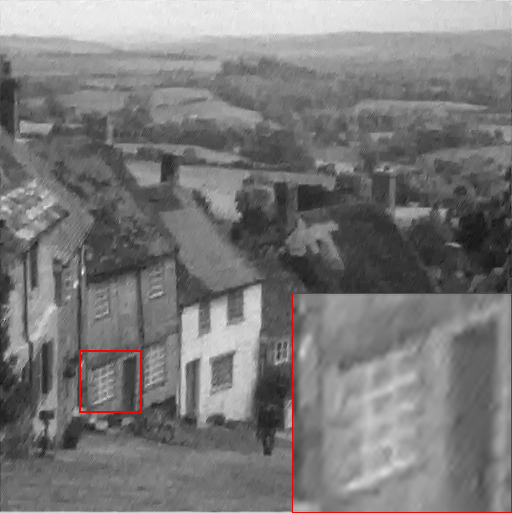}&
\includegraphics[width=0.23\linewidth]{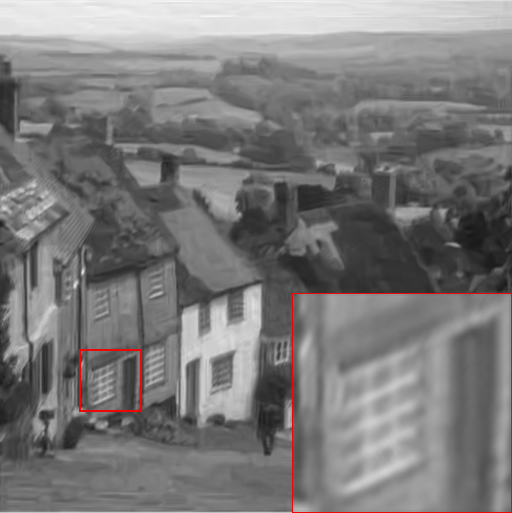}&
\includegraphics[width=0.23\linewidth]{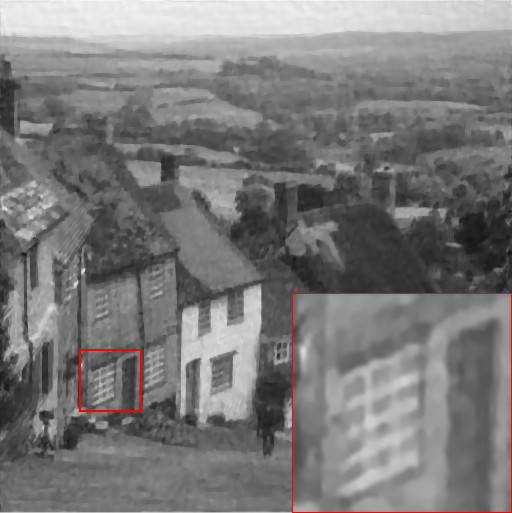}\\
\small (e) TRL2 ~\cite{2018Ag}&
\small (f) SOCF ~\cite{2020Ani}&
\small (g) BM3D ~\cite{2008Image}&
\small (h) Ours\\
\small 27.70/0.692&
\small 28.36/0.712&
\small 28.53/0.717&
\small 28.56/0.722\\
 \end{tabular}
\caption{ Deblurring result of MB(20, 60) and $\sigma=3$ for gray image "Hill" with zoomed areas and PSNR values and SSIM values. }
\label{fig:MBg}
\centering
\end{figure}

\begin{figure}[!ht]
\centering
%\renewcommand{\arraystretch}{0.2}\addtolength{\tabcolsep}{-5pt} \vskip3mm
%\backslash\footsize{1pt}{\baselineskip}\selectfont
 \begin{tabular}{c@{\hskip 2.5pt}c@{\hskip 2.5pt}c@{\hskip 2.5pt}cc}
    \footnotesize
\includegraphics[width=0.23\linewidth]{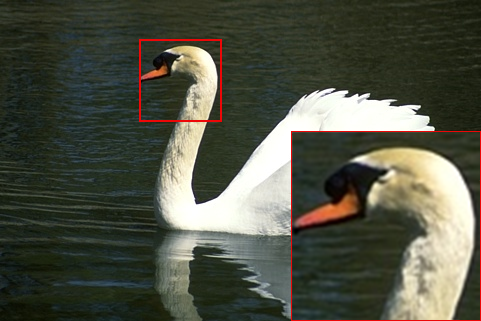}&
\includegraphics[width=0.23\linewidth]{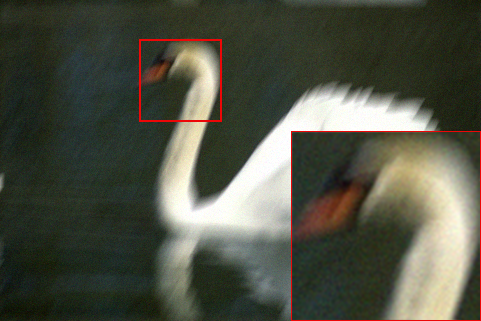}&
\includegraphics[width=0.23\linewidth]{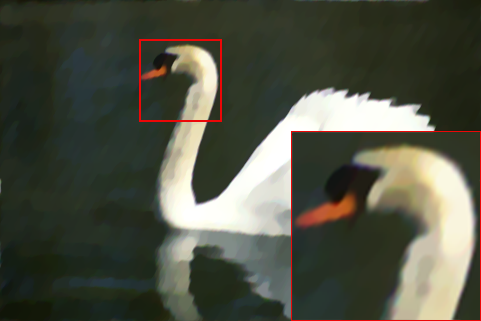}&
\includegraphics[width=0.23\linewidth]{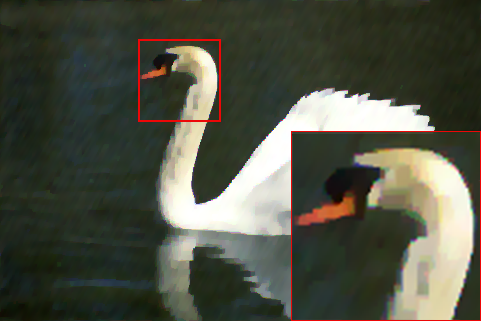}&\\
\small (a) Original image&
\small (b) Degraded image&
\small (c) TV ~\cite{1992Nonlinear}&
\small (d) DCA ~\cite{2015Aw}\\
&
\small 22.48/0.623&
\small 26.79/0.782&
\small 27.36/0.793\\
\includegraphics[width=0.23\linewidth]{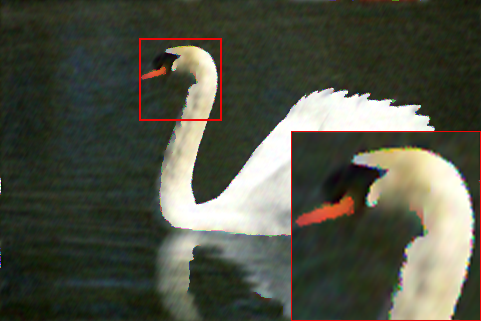}&
\includegraphics[width=0.23\linewidth]{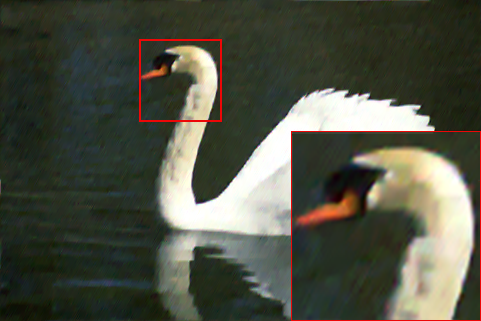}&
\includegraphics[width=0.23\linewidth]{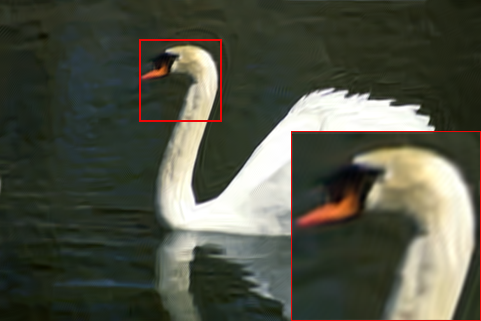}&
\includegraphics[width=0.23\linewidth]{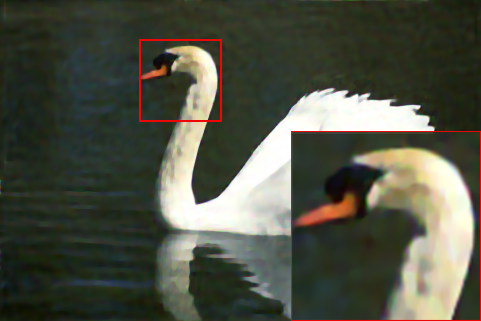}\\
\small (e) TRL2 ~\cite{2018Ag}&
\small (f) SOCF ~\cite{2020Ani}&
\small (g) BM3D ~\cite{2008Image}&
\small (h) Ours\\
\small 26.61/0.683&
\small 27.98/0.792&
\small 28.03/0.808&
\small 28.57/0.797
 \end{tabular}
\caption{ Deblurring result of MB(20, 60) and $\sigma=5$ for color image "Duck" with zoomed areas and PSNR values and SSIM values. }
\label{fig:MBc}
\centering
\end{figure}

\begin{figure}[!ht]
\centering
%\renewcommand{\arraystretch}{0.2}\addtolength{\tabcolsep}{-5pt} \vskip3mm
%\backslash\footsize{1pt}{\baselineskip}\selectfont
 \begin{tabular}{c@{\hskip 2.5pt}c@{\hskip 2.5pt}c@{\hskip 2.5pt}cc}
    \footnotesize
\includegraphics[width=0.23\linewidth]{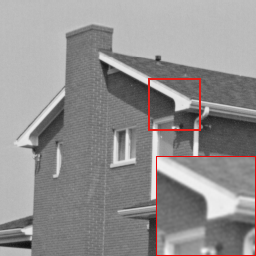}&
\includegraphics[width=0.23\linewidth]{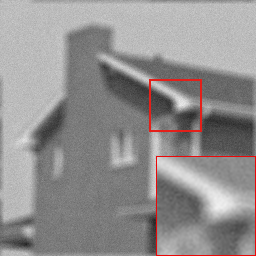}&
\includegraphics[width=0.23\linewidth]{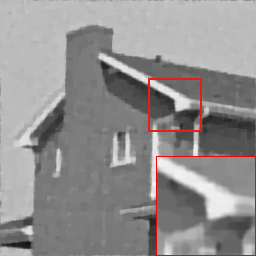}&
\includegraphics[width=0.23\linewidth]{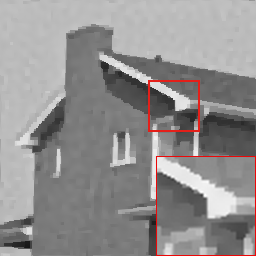}&\\
\small (a) Original image&
\small (b) Degraded image&
\small (c) TV ~\cite{1992Nonlinear}&
\small (d) DCA ~\cite{2015Aw}\\
&
\small 23.70/0.543&
\small 29.60/0.798&
\small 28.76/0.802\\
\includegraphics[width=0.23\linewidth]{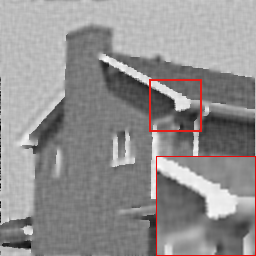}&
\includegraphics[width=0.23\linewidth]{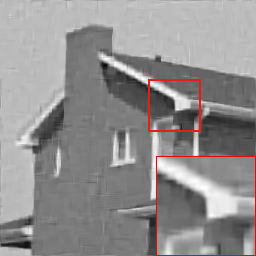}&
\includegraphics[width=0.23\linewidth]{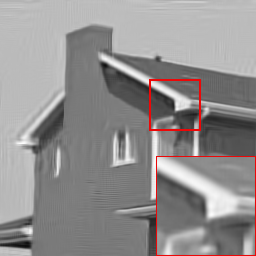}&
\includegraphics[width=0.23\linewidth]{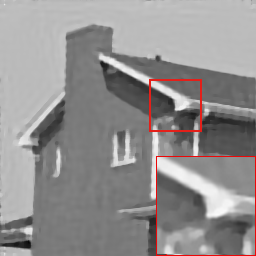}\\
\small (e) TRL2 ~\cite{2018Ag}&
\small (f) SOCF ~\cite{2020Ani}&
\small (g) BM3D ~\cite{2008Image}&
\small (h) Ours\\
\small 26.80/0.592&
\small 29.75/0.796&
\small 30.33/0.808&
\small 30.49/0.817
 \end{tabular}
\caption{ Deblurring result of AB(9, 9) and $\sigma=5$ for gray image "House" with zoomed areas and PSNR values and SSIM values. }
\label{fig:ABg}
\centering
\end{figure}

\begin{figure}[!ht]
\centering
%\renewcommand{\arraystretch}{0.2}\addtolength{\tabcolsep}{-5pt} \vskip3mm
%\backslash\footsize{1pt}{\baselineskip}\selectfont
 \begin{tabular}{c@{\hskip 2.5pt}c@{\hskip 2.5pt}c@{\hskip 2.5pt}cc}
    \footnotesize
\includegraphics[width=0.23\linewidth]{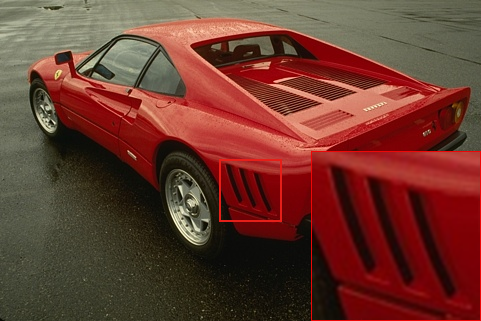}&
\includegraphics[width=0.23\linewidth]{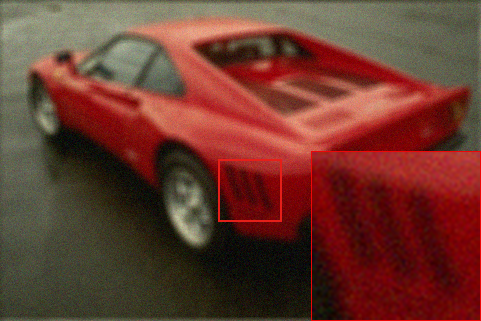}&
\includegraphics[width=0.23\linewidth]{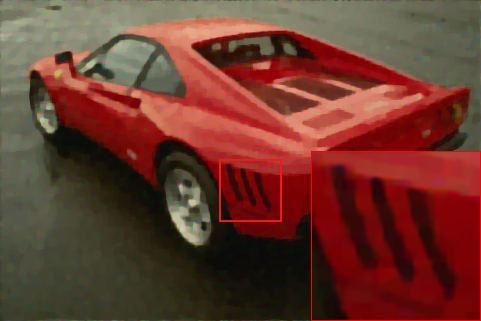}&
\includegraphics[width=0.23\linewidth]{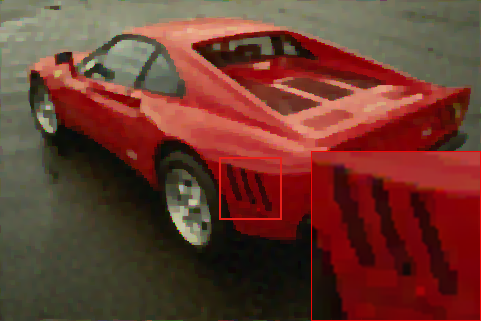}&\\
\small (a) Original image&
\small (b) Degraded image&
\small (c) TV ~\cite{1992Nonlinear}&
\small (d) DCA ~\cite{2015Aw}\\
&
\small 23.10/0.685&
\small 26.14/0.834&
\small 25.56/0.825\\
\includegraphics[width=0.23\linewidth]{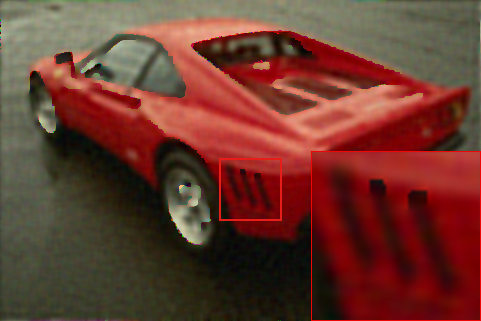}&
\includegraphics[width=0.23\linewidth]{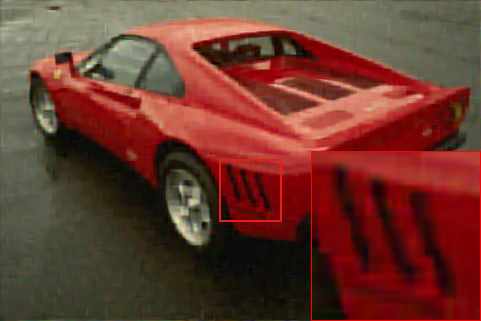}&
\includegraphics[width=0.23\linewidth]{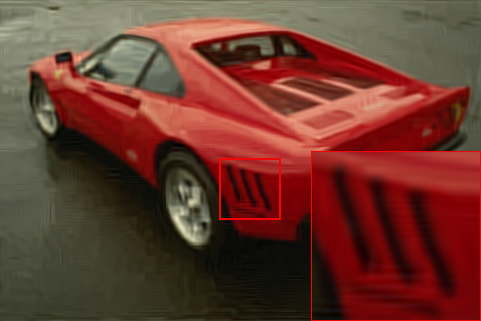}&
\includegraphics[width=0.23\linewidth]{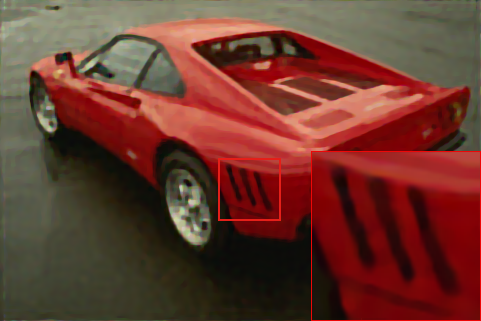}\\
\small (e) TRL2 ~\cite{2018Ag}&
\small (f) SOCF ~\cite{2020Ani}&
\small (g) BM3D ~\cite{2008Image}&
\small (h) Ours\\
\small 25.18/0.799&
\small 25.95/0.823&
\small 26.24/0.839&
\small 26.54/0.842\\
 \end{tabular}
\caption{ Deblurring result of AB(9, 9) and $\sigma=8$ for color image "Car" with zoomed areas and PSNR values and SSIM values. }
\label{fig:ABc}
\centering
\end{figure}

%\textcolor{red}{
In order to further demonstrate the superiority of our algorithm on image restoration more intuitively,  
we compare the visual quality of restored images of our proposed method and other state-of-the-art methods. 
%we present visual results of image deblurring.  %In order to, 
We also zoom in key parts of the image for better illustrating.  % the visual quality.  
%In particular,  we consider Gaussian blur $\mathrm{GB(9, 5)}/\sigma=5$,  Motion blur  $\mathrm{MB(20, 60)}/\sigma=5$ and Average blur $\mathrm{AB(9, 9)}/\sigma=3$ for Shepp-Logan,  Cameraman and Duck. 
%\textcolor{red}{In particular, when presenting the effects of a grey-scale image, we choose $\mathrm{GB(9, 5)}/\sigma=8$ for Shepp--logan, $\mathrm{MB(20, 60)}/\sigma=3$ for hill and $\mathrm{AB(9, 9)}/\sigma=5$ for house.  When showing the effect of the coloured figure, we choose $\mathrm{GB(9, 5)}/\sigma=10$ for the plate, $\mathrm{MB(20, 60)}/\sigma=5$ duck and $\mathrm{AB(9, 9)}/\sigma=8$ for the car.  The small box with the thicker red edge in the diagram is the local area that we need to zoom in on, and the large box with the thinner red edge in the bottom right corner is the local area that we need to zoom in on appropriately. 
%}
Figures $\ref{fig:GBg}-\ref{fig:ABc}$ show that our method yields the best image quality in terms of removing noise,  preserving edges,  and maintaining image sharpness. 
Looking carefully at the details in the figures,  we can see that TV ~\cite{1992Nonlinear} oversmoothes images,  TRL2 ~\cite{2018Ag} introduces staircase effect,  DCA ~\cite{2015Aw} sharpens edges but loses some detail information,  there exits the ringing phenomenon by  SOCF ~\cite{2020Ani} and BM3D model ~\cite{2008Image} introduces some artifacts.  
%}
%\textcolor{red}{The image restored from the BM3D model ~\cite{2008Image} will have certain artifacts.  }
%Figures \ref{fig:GBg} $-$ \ref{fig:ABc} show that our algorithm gets even greater advantages. 
%Methods TV ~\cite{1992Nonlinear},  DCA ~\cite{2015Aw} and SOCF ~\cite{2020Ani} have the same problem with $\mathrm{GB(9, 5)}/\sigma=5$, then BM3D~\cite{2008Image} and TRL2 ~\cite{2018Ag} lefts excessive residual noise,while the proposed method always maintains good visual quality. 

%Tables \ref{tab:table e},  \ref{tab:table g} and \ref{tab:table AB} confirm that our method achieves the best results in PSNR and SSIM in most cases. 
 Table \ref{tab:table 2} lists the computation time on a desktop  (Intel(R) Core(TM) i5-8250 CPU @1.60 GHz),  which reveal that our method is faster than DCA ~\cite{2015Aw},  but is  slower than TV ~\cite{1992Nonlinear}, BM3D ~\cite{2008Image}, TRL2 \cite{2018Ag} and SOCF \cite{2020Ani}. Acceleration will be considered in the future work.

In brief,  our algorithm (EAHR) has competitive performance in sharpening the edges and removing the noise,  which outperforms other mainstream methods both in PSNR/SSIM values and visual quality.

\section{Conclusion}

For an entire image,  the spatially fixed regularization parameters can not perform well for both edges and smooth areas. The larger parameters  are favorable to reduce the noise in the smooth area,  while incurs  blur to edges.  The small parameters enable regularization-based algorithms to sharpen the edges,  but the denoising may not be sufficient. % cause residual noise.}

In this paper,  we have presented an automated spatially dependent regularization parameter selection framework
 for restoring an image from a noisy and blur image.
An edge detector with high  robustness to noise is used to detect the edges of the image and then generates an edge information matrix.
According to this matrix, the automated spatially dependent parameters for regularization  are given in binarization.
With the help of these parameters, the regularization algorithm performs outstandingly  at both edges and smooth areas.
Once automated spatially dependent parameters are fixed, the proposed model is convex  and therefore can be solved by  sPADMM with a linear-rate convergence rate.
Extensive experiments on different types of blurring kernels and different levels of Gaussian noise have been conducted to show that our approach is robust and outperforms other state-of-the-art deblurring methods.
In addition, the proposed model not only
%could eliminate
effectively  overcomes the false edges and staircase effects % generated in the TV denoising process,
but also makes up for the insufficiency of the harmonic model's diffusion  in all directions and protects the edge well to restore the internal smooth area.
Due to the limited space,  we only display the experimental results in three cases: Gaussian blur $\mathrm{GB(9, 5)}/\sigma=5$,  Motion blur  $\mathrm{MB(20, 60)}/\sigma=5$ and Average blur  $\mathrm{AB(9, 9)}/\sigma=3$.
The future work will include:
\begin{enumerate}
  \item We will accelerate our algorithm to reduce the recovery time.
  \item To  remain more image details, we will examine an  edge detector with higher accuracy and robustness.
  Moreover, the automated spatially dependent parameters will be considered according to the image texture.
  \item Our model is now only suitable for non-blind deblurring,  and we will extend it to blind deblurring task.
\end{enumerate}

\section*{Acknowledgement}
The work has been supported by the National Natural Science Foundation of China (Grants Nos. 12061052), Natural Science Fund of Inner Mongolia Autonomous Region (Grant No. 2020MS01002), the China Scholarship Council for a one year visiting at Ecole Normale Sup\'{e}rieure Paris-Saclay (No. 201806810001).  The authors would like to thank Professor Guoqing Chen for providing us some useful suggestions and this work is also partly supported by his project (“111 project”of higher education talent training in Inner Mongolia Autonomous Region).

\section*{Appendix. Convergence proof}

 To make the paper self-contained, we establish the global convergence result of algorithm 1.

Let $F(\mathbf{u})=\frac{\mu}{2}\|\mathbf{Au}-\mathbf{f}\|_{2}^{2}$,  and  $G(\mathbf{k})=\|\mathbf{k}\|_{1, \mathbf{\alpha}_{1}}+\frac{1}{2}\|\mathbf{k}\|_{2, \mathbf{\alpha}_{2}}^{2}$. If $(\bar{\mathbf{u}}, \bar{\mathbf{k}})$ is an optimal solution of (\ref{yue}) if and only if there exists Lagrange multiplier $\bar{\mathbf{\lambda}}$ such that
\begin{equation*}
\cases{0=\nabla F(\bar{\mathbf{u}})+\nabla^{\top}\bar{\mathbf{\lambda}},\\
0\in\partial G(\bar{\mathbf{k}})-\bar{\mathbf{\lambda}},\\
0=\mathbf{\nabla}\bar{\mathbf{u}}-\bar{\mathbf{k}},\\}
\end{equation*}
where $\nabla F=\mu\mathbf{A}^{\top}(\mathbf{Au}-\mathbf{f}), $ and $\partial G$ is the subdifferential of $G$. From the KKT conditions,  we know that if $\{(\mathbf{u}, \mathbf{k}, \mathbf{\lambda})\}$  is the KKT point of (\ref{yue}),  then
\begin{equation}\label{21}
\cases{0=\nabla F(\mathbf{u})+\nabla^{\top}\mathbf{\lambda}, \\
0\in\partial G(\mathbf{k})-\mathbf{\lambda}, \\
0=\nabla \mathbf{u}-\mathbf{k}.\\}
\end{equation}
Since the subdifferential mapping of the closed convex function is maximal monotone,  there
exist self-adjoint and positive semidefinite operators $\sum_{\mathbf{F}},  \sum_{\mathbf{G}}$ such that for all $\mathbf{u},  \hat{\mathbf{u}}\in dom(F),  \omega\in\nabla F(\mathbf{u})$ and $\hat{\omega}\in\nabla F(\hat{\mathbf{u}})$,
\begin{equation}
\langle\omega-\hat{\omega},  \mathbf{u}-\hat{\mathbf{u}}\rangle\geq\|\mathbf{u}-\hat{\mathbf{u}}\|_{\sum_{\mathbf{F}}}^{2},
\label{3}
\end{equation}
for all $\mathbf{k}, \hat{\mathbf{k}}\in dom (G), x\in\partial G(\mathbf{k})$ and $\hat x\in\partial G(\hat{\mathbf{k}})$,
\begin{equation}
\langle x-\hat x, \mathbf{k}-\hat{\mathbf{k}}\rangle\geq\|\mathbf{k}-\hat{\mathbf{k}}\|_{\sum_{\mathbf{G}}}^{2}.
\label{4}
\end{equation}
It is obtained by our algorithm that
\begin{equation}\label{6}
\cases{0=\nabla F(\mathbf{u}^{n+1})+\nabla^{\top}(\beta(\nabla \mathbf{u}^{n+1}-\mathbf{k}^{n})+\mathbf{\lambda}^{n})+\mathbf{S_{1}}(\mathbf{u}^{n+1}-\mathbf{u}^{n}), \\
0\in\partial G(\mathbf{k}^{n+1})-(\beta(\nabla \mathbf{u}^{n+1}-\mathbf{k}^{n+1})+\mathbf{\lambda}^{n})+\mathbf{S_{2}}(\mathbf{k}^{n+1}-\mathbf{k}^{n}), \\
0=\nabla \mathbf{u}^{n+1}-\mathbf{k}^{n+1}-(\beta\eta)^{-1}(\mathbf{\lambda}^{n+1}-\mathbf{\lambda}^{n}).\\}
\end{equation}
Let $\epsilon(\mathbf{u}, \mathbf{k})=\nabla \mathbf{u}-\mathbf{k}$. Denote $\mathbf{u}_{e}^{n}=\mathbf{u}^{n}-\bar{\mathbf{u}}$,  similarly $\mathbf{k}_{e}^{n}, \mathbf{\lambda}_{e}^{n}$,
\begin{equation*}
\cases{w^{n+1}=\mathbf{\lambda}^{n+1}+(1-\eta)\beta\epsilon(\mathbf{u}^{n+1}, \mathbf{k}^{n+1})+\beta(\mathbf{k}^{n+1}-\mathbf{k}^{n}), \\
x^{n+1}=\mathbf{\lambda}^{n+1}+(1-\eta)\beta\epsilon(\mathbf{u}^{n+1}, \mathbf{k}^{n+1}).\\}
\end{equation*}
Therefore,
\begin{equation*}
%\eqalign{ &\|{\mathbf{u}_{e}^{n+1}}\|_{\sum_{F}}^{2}\\
%& \leq\langle w^{n+1}-\overline{w}, \mathbf{u}^{n+1}-\overline{\mathbf{u}}\rangle\cr
%&=\langle \nabla^{\top}(\bar{\mathbf{\lambda}}-\mathbf{\lambda}^{n}) + \beta(\nabla^{\top}
%\mathbf{k}^{n}-\mathbf{\triangle} \mathbf{u}^{n+1})%-\beta\mathbf{\triangle} \mathbf{u}^{n+1}+\beta\nabla^{\top}
%\mathbf{k}^{n}
%-\mathbf{S_{1}}(\mathbf{u}^{n+1}-\mathbf{u}^{n}), \mathbf{u}_{e}^{n+1}\rangle,\\
%\fl
\eqalign{\epsilon( \mathbf{u}_{e}^{n+1}, \mathbf{k}_{e}^{n+1})
%&=\nabla (\mathbf{u}^{n+1}-\overline{\mathbf{u}})-(\mathbf{k}^{n+1}-\overline{\mathbf{k}})\\
%&=\nabla \mathbf{u}^{n+1}-\nabla\overline{\mathbf{u}}-\mathbf{k}^{n+1}+\overline{\mathbf{k}}\quad\quad{\color{blue}{(note:\bar{k}=\nabla\bar{u})}}\\
&=\nabla \mathbf{u}^{n+1}-\mathbf{k}^{n+1}=\epsilon(\mathbf{u}^{n+1}, \mathbf{k}^{n+1})\\
&=(\beta\eta)^{-1}(\mathbf{\lambda}_{e}^{n+1}-\mathbf{\lambda}_{e}^{n})=(\beta\eta)^{-1}(\mathbf{\lambda}^{n+1}-\mathbf{\lambda}^{n}).}
%&=(\beta\eta)^{-1}(\mathbf{\lambda}^{n+1}-\mathbf{\lambda}^{n})
\end{equation*}
Noting (\ref{21}), (\ref{3}), (\ref{4}) and (\ref{6}), we find
\begin{equation}\label{7}
%\qquad \quad \fl 
\eqalign{ &\|{\mathbf{u}_{e}^{n+1}}\|_{\sum_{F}}^{2}\\
& \leq\langle w^{n+1}-\overline{w}, \mathbf{u}^{n+1}-\overline{\mathbf{u}}\rangle\cr
&=\langle \nabla^{\top}(\bar{\mathbf{\lambda}}-\mathbf{\lambda}^{n}) + \beta(\nabla^{\top}
\mathbf{k}^{n}-\mathbf{\triangle} \mathbf{u}^{n+1})%-\beta\mathbf{\triangle} \mathbf{u}^{n+1}+\beta\nabla^{\top}
%\mathbf{k}^{n}
-\mathbf{S_{1}}(\mathbf{u}^{n+1}-\mathbf{u}^{n}), \mathbf{u}_{e}^{n+1}\rangle,}
%&=\langle-\nabla^{\top}\mathbf{\lambda}^{n}-\beta\mathbf{\triangle} \mathbf{u}^{n+1}+\beta\nabla^{\top}
%\mathbf{k}^{n}-\mathbf{S_{1}}(\mathbf{u}^{n+1}-\mathbf{u}^{n})+\nabla^{\top}\bar{\mathbf{\lambda}}, \mathbf{u}_{e}^{n+1}\rangle,}
\end{equation}
\begin{equation}\label{8}
%\qquad \quad \fl 
\eqalign{ &\|{\mathbf{k}_{e}^{n+1}}\|_{\sum_{\mathbf{G}}}^{2}
\\& \leq\langle x^{n+1}-\overline{x}, \mathbf{k}^{n+1}-\overline{\mathbf{k}}\rangle\\
%\quad\quad\quad{\color{blue}(note:x^{n+1}\in\partial G(k^{n+1}), \overline{x}\in\partial G(\overline{k}))}\\
&=\langle \mathbf{\lambda}^{n}-\bar{\mathbf{\lambda}}+\beta(\nabla \mathbf{u}^{n+1}-\mathbf{k}^{n+1})-\mathbf{S_{2}}(\mathbf{k}^{n+1}-\mathbf{k}^{n})
, \mathbf{k}_{e}^{n+1}\rangle.}
\end{equation}
By calculation,  we get
\begin{equation*}
\eqalign{ -\nabla^{\top}w^{n+1}&=-\nabla^{\top}[\mathbf{\lambda}^{n+1}+(1-\eta)\beta\epsilon( \mathbf{u}^{n+1}, \mathbf{k}^{n+1})+\beta(\mathbf{k}^{n+1}-\mathbf{k}^{n})]\\
%&=-\nabla^{\top}[\mathbf{\lambda}^{n+1}+(1-\eta)\beta(\eta\beta)^{-1}(\mathbf{\lambda}^{n+1}-\mathbf{\lambda}^{n})
%+\beta(\mathbf{k}^{n+1}-\mathbf{k}^{n})]\\
%&=-\nabla^{\top}[\mathbf{\lambda}^{n+1}+(1-\eta)\eta^{-1}(\mathbf{\lambda}^{n+1}-\mathbf{\lambda}^{n})+\beta(\mathbf{k}^{n+1}-\mathbf{k}^{n})]\\
%&=-\nabla^{\top}[\mathbf{\lambda}^{n+1}+(\eta^{-1}-1)(\mathbf{\lambda}^{n+1}-\mathbf{\lambda}^{n})+\beta(\mathbf{k}^{n+1}-\mathbf{k}^{n})]\\
%&=-\nabla^{\top}[\mathbf{\lambda}^{n+1}+\eta^{-1}(\mathbf{\lambda}^{n+1}-\mathbf{\lambda}^{n})-\mathbf{\lambda}^{n+1}+
%\mathbf{\lambda}^{n}+\beta(\mathbf{k}^{n+1}-\mathbf{k}^{n})]\\
%&=-\nabla^{\top}(\mathbf{\lambda}^{n}+\beta\nabla \mathbf{u}^{n+1}-\beta \mathbf{k}^{n+1}+\beta \mathbf{k}^{n+1}-\beta \mathbf{k}^{n})\\
&=-\nabla^{\top}\mathbf{\lambda}^{n}-\beta\nabla^{\top}\nabla \mathbf{u}^{n+1}+\beta\nabla^{\top} \mathbf{k}^{n},}
\end{equation*}
\begin{equation*}
\qquad ~\eqalign{x^{n+1}&=\mathbf{\lambda}^{n+1}+(1-\eta)\beta\epsilon(\mathbf{u}^{n+1}, \mathbf{k}^{n+1})\\
%&=\mathbf{\lambda}^{n+1}+(1-\eta)\beta(\eta\beta)^{-1}(\mathbf{\lambda}^{n+1}-\mathbf{\lambda}^{n})\\
%&=\mathbf{\lambda}^{n+1}+(\eta^{-1}-1)(\mathbf{\lambda}^{n+1}-\mathbf{\lambda}^{n})\\
%&=\mathbf{\lambda}^{n+1}+\eta^{-1}(\mathbf{\lambda}^{n+1}-\mathbf{\lambda}^{n})-\mathbf{\lambda}^{n+1}+\mathbf{\lambda}^{n}\\
%&=\mathbf{\lambda}^{n}+\eta^{-1}(\eta\beta)(\nabla \mathbf{u}^{n+1}-\mathbf{k}^{n+1})\\
&=\mathbf{\lambda}^{n}+\beta(\nabla \mathbf{u}^{n+1}-\mathbf{k}^{n+1}).}
\end{equation*}
Thus,  (\ref{7}) is converted to (\ref{9}) and (\ref{8}) becomes (\ref{10}),
\begin{equation}\label{9}
\|{\mathbf{u}_{e}^{n+1}}\|_{\sum_{\mathbf{F}}}^{2}\leq\langle-\nabla^{\top}w^{n+1}-\mathbf{S_{1}}(\mathbf{u}^{n+1}
-\mathbf{u}^{n})+\nabla^{\top}\overline{\mathbf{\lambda}}, \mathbf{u}_{e}^{n+1}\rangle,
\end{equation}
\begin{equation}\label{10}
\|{\mathbf{k}_{e}^{n+1}}\|_{\sum_{\mathbf{G}}}^{2}\leq\langle x^{n+1}-\mathbf{S_{2}}(\mathbf{k}^{n+1}-\mathbf{k}^{n})-\overline{\mathbf{\lambda}}, \mathbf{k}_{e}^{n+1}\rangle.
\end{equation}
Then through (\ref{9}) and (\ref{10}),  we obtain
\begin{equation}\label{11}
%\quad \fl 
\eqalign{ \|&{\mathbf{u}_{e}^{n+1}}\|_{\sum_{\mathbf{F}}}^{2}+\|{\mathbf{k}_{e}^{n+1}}\|_{\sum_{\mathbf{G}}}^{2}\\
&\leq(\beta\eta)^{-1}\langle\mathbf{\lambda}_{e}^{n+1}, \mathbf{\lambda}_{e}^{n}-\mathbf{\lambda}_{e}^{n+1}\rangle
-\beta\langle\mathbf{k}^{n+1}-\mathbf{k}^{n}, \epsilon(\mathbf{u}^{n+1}, \mathbf{k}^{n+1})\rangle\\
&-\beta\langle\mathbf{k}^{n+1}-\mathbf{k}^{n}, \mathbf k_{e}^{n+1}\rangle
-\beta(1-\eta)\|\epsilon(\mathbf{u}^{n+1}, \mathbf{k}^{n+1})\|^{2}_{2}\\
&- \langle \mathbf{S_{1}}(\mathbf{u}^{n+1}-\mathbf{u}^{n}), \mathbf{u}_{e}^{n+1}\rangle-\langle \mathbf{S_{2}}(\mathbf{k}^{n+1}-\mathbf{k}^{n}), \mathbf{k}_{e}^{n+1}\rangle.\\}
\end{equation}
Next, we shall estimate the term $\beta\langle \mathbf{k}^{n+1}-\mathbf{k}^{n}, \epsilon (\mathbf{u}^{n+1}, \mathbf{k}^{n+1})\rangle$. It follows from equation (\ref{6}) that
 \begin{equation*}
 x^{n+1}-\mathbf{S_{2}}(\mathbf{k}^{n+1}-\mathbf{k}^{n})\in\partial G(\mathbf{k}^{n+1}), ~~
x^{n}-\mathbf{S_{2}(}\mathbf{k}^{n}-\mathbf{k}^{n-1})\in\partial G(\mathbf{k}^{n}). 
 \end{equation*}
In addition,  by the the maximal monotonic property of $\partial G(.)$,  we have
\begin{equation*}
\eqalign{ &\langle \mathbf{k}^{n+1}-\mathbf{k}^n, x^{n+1}-x^n\rangle\\
%=&\langle\mathbf{k}^{n+1}-\mathbf{k}^n, x^{n+1}-\mathbf{S_{2}}(\mathbf{k}^{n+1}-\mathbf{k}^n)+\mathbf{S_{2}}(\mathbf{k}^{n+1}-\mathbf{k}^n)
%-x^n+\mathbf{S_{2}}(\mathbf{k}^{n}-\mathbf{k}^{n-1})-\mathbf{S_{2}}(\mathbf{k}^{n}-\mathbf{k}^{n-1})\rangle\\
&=\langle\mathbf{k}^{n+1}-\mathbf{k}^n, x^{n+1}-\mathbf{S_{2}}(\mathbf{k}^{n+1}-\mathbf{k}^n)-(x^n-\mathbf{S_{2}}(\mathbf{k}^{n}-\mathbf{k}^{n-1}))\rangle\\
%&{\color{blue}{(note:x^{n+1}-\mathbf{S_{2}}(\mathbf{k}^{n+1}-\mathbf{k}^n)\in\partial G(\mathbf{k}^{n+1}), x^n-\mathbf{S_{2}}(\mathbf{k}^{n}-\mathbf{k}^{n-1}\in\partial G(\mathbf{k}^n))}}\\
&\quad +\langle\mathbf{k}^{n+1}-\mathbf{k}^n, \mathbf{S_{2}}(\mathbf{k}^{n+1}-\mathbf{k}^n)\rangle
-\langle\mathbf{k}^{n+1}-\mathbf{k}^n, \mathbf{S_{2}}(\mathbf{k}^n-\mathbf{k}^{n-1})\rangle\\
%&{\color{blue}{(note:\langle\mathbf{k}^{n+1}-\mathbf{k}^n, \mathbf{S_{2}}(\mathbf{k}^{n+1}-\mathbf{k}^n)\rangle=\|\mathbf{k}^{n+1}-\mathbf{k}^n\|^{2}_\mathbf{{S_{2}}})}}\\
&\geq\|\mathbf{k}^{n+1}-\mathbf{k}^n\|^{2}_\mathbf{{S_{2}}}+\|\mathbf{k}^{n+1}-\mathbf{k}^n\|^{2}_\mathbf{{S_{2}}}
-\langle \mathbf{k}^{n+1}-\mathbf{k}^n, \mathbf{S_{2}}(\mathbf{k}^n-\mathbf{k}^{n-1})\rangle\\
&\geq\|\mathbf{k}^{n+1}-\mathbf{k}^n\|^{2}_\mathbf{{S_{2}}}-\langle \mathbf{k}^{n+1}-\mathbf{k}^n, \mathbf{S_{2}}(\mathbf{k}^n-\mathbf{k}^{n-1})\rangle.}
\end{equation*}
Let $\alpha_{n+1}=-(1-\eta)\beta\langle \mathbf{k}^{n+1}-\mathbf{k}^n, \epsilon(\mathbf{u}^{n}, \mathbf{k}^n)\rangle$,  then
\begin{equation}\label{12}
\eqalign{ &-\beta\langle \mathbf{k}^{n+1}-\mathbf{k}^n, \epsilon(\mathbf{u}^{n+1}, \mathbf{k}^{n+1})\rangle\\
%=&-(1-\eta)\beta\langle \mathbf{k}^{n+1}-\mathbf{k}^n, \epsilon(\mathbf{u}^{n+1}, \mathbf{k}^{n+1})\rangle-\beta\eta\langle \mathbf{k}^{n+1}-\mathbf{k}^n, \epsilon(\mathbf{u}^{n+1}, \mathbf{k}^{n+1})\rangle\\
%=&-(1-\eta)\beta\langle \mathbf{k}^{n+1}-\mathbf{k}^n, \epsilon(\mathbf{u}^{n+1}, \mathbf{k}^{n+1})\rangle-\beta\eta\langle \mathbf{k}^{n+1}-\mathbf{k}^n, (\eta\beta)^{-1}(\mathbf{\lambda}^{n+1}-\mathbf{\lambda}^n)\rangle\\
&=-(1-\eta)\beta\langle \mathbf{k}^{n+1}-\mathbf{k}^n, \epsilon(\mathbf{u}^{n+1}, \mathbf{k}^{n+1})\rangle - \langle \mathbf{k}^{n+1}-\mathbf{k}^n, \mathbf{\lambda}^{n+1}-\mathbf{\lambda}^{n}\rangle\\
&=-(1-\eta)\beta\langle \mathbf{k}^{n+1}-\mathbf{k}^n, \epsilon(\mathbf{u}^{n+1}, \mathbf{k}^{n+1})\rangle\\
&\quad -\langle\mathbf{k}^{n+1}-\mathbf{k}^n, {{x}^{n+1}-{x}^n-(1-\eta)\beta\epsilon(\mathbf{u}^{n+1}, \mathbf{k}^{n+1})%\.
%\\& \quad \.
+(1-\eta)\beta\epsilon(\mathbf{u}^{n}, \mathbf{k}^{n})\rangle}\\
%&{\color{red}{(note:x^{n+1}-x^{n}=\mathbf{\lambda}^{n+1}-\mathbf{\lambda}^n+(1-\eta)\beta\epsilon(\mathbf{u}^{n+1}, \mathbf{k}^{n+1})-(1-\eta)\beta\epsilon(\mathbf{u}^{n}, \mathbf{k}^{n}))}}\\
%=&-\beta(1-\eta)\langle \mathbf{k}^{n+1}-\mathbf{k}^n, \epsilon(\mathbf{u}^{n+1}, \mathbf{k}^{n+1})\rangle
%-\langle \mathbf{k}^{n+1}-\mathbf{k}^n, {x}^{n+1}-{x}^n\rangle\\
%&+\beta(1-\eta)\langle \mathbf{k}^{n+1}-\mathbf{k}^{n}, \epsilon(\mathbf{u}^{n+1}, \mathbf{k}^{n+1})\rangle
%-\beta(1-\eta)\langle \mathbf{k}^{n+1}-\mathbf{k}^n, \epsilon(\mathbf{u}^{n}, \mathbf{k}^{n})\rangle\\
%&{\color{blue}{(note:\alpha_{n+1}=-\beta(1-\eta)\langle \mathbf{k}^{n+1}-\mathbf{k}^n, \epsilon(\mathbf{u}^{n}, \mathbf{k}^{n})\rangle)}}\\
&=\alpha_{n+1}+\langle \mathbf{k}^{n+1}-\mathbf{k}^n, {x}^{n}-{x}^{n+1}\rangle\\
&\leq\alpha_{n+1}-\|\mathbf{k}^{n+1}-\mathbf{k}^n\|^{2}_\mathbf{{\mathbf{S_{2}}}}+\langle \mathbf{k}^{n+1}-\mathbf{k}^n, \mathbf{\mathbf{S_{2}}}(\mathbf{k}^n-\mathbf{k}^{n-1})\rangle\\
&\leq\alpha_{n+1}-\|\mathbf{k}^{n+1}-\mathbf{k}^n\|^{2}_\mathbf{\mathbf{{S_{2}}}}+\|\mathbf{k}^{n+1}-
\mathbf{k}^n\|_\mathbf{{\mathbf{S_{2}}}}\|\mathbf{S_{2}}(\mathbf{k}^n
%\\&
-\mathbf{k}^{n-1})\|_\mathbf{{S_{2}}^{-1}}\\
&\leq\alpha_{n+1}-\|\mathbf{k}^{n+1}-\mathbf{k}^n\|^{2}_{\mathbf{S_{2}}}+\frac{1}{2}\|\mathbf{k}^{n+1}-
\mathbf{k}^n\|^{2}_{\mathbf{S_{2}}}+\frac{1}{2}\|\mathbf{k}^{n}-\mathbf{k}^{n-1}
\|^{2}_{\mathbf{S_{2}}}\\
&={\alpha_{n+1}-\frac{1}{2}\|\mathbf{k}^{n+1}-\mathbf{k}^n\|^{2}_{\mathbf{S_{2}}}
+\frac{1}{2}\|\mathbf{k}^{n}-\mathbf{k}^{n-1}\|^{2}_{\mathbf{S_{2}}}}.}
\end{equation}
Since $\mathbf{\lambda}^{n+1}=\mathbf{\lambda}^{n}+(\beta\eta)\epsilon(\mathbf{u}^{n+1}, \mathbf{k}^{n+1})$,
it follows from (\ref{11}) and (\ref{12}) that
\begin{equation}\label{13}
\eqalign{ 2&\|{\mathbf{u}_{e}^{n+1}}\|_{\sum_{\mathbf{F}}}^{2}+2\|{\mathbf{k}_{e}^{n+1}}\|_{\sum_{\mathbf{G}}}^{2}\\
\leq & 2(\beta\eta)^{-1}\langle\mathbf{\lambda}_{e}^{n+1}, \mathbf{\lambda}_{e}^{n}-\mathbf{\lambda}_{e}^{n+1}\rangle
-2\beta\langle\mathbf{k}^{n+1}-\mathbf{k}^{n}, \epsilon(\mathbf{u}^{n+1}, \mathbf{k}^{n+1})\rangle\\
&-2\beta\langle\mathbf{k}^{n+1}-\mathbf{k}^{n}, \mathbf k_{e}^{n+1}\rangle
-2\beta(1-\eta)\|\epsilon(\mathbf{u}^{n+1}, \mathbf{k}^{n+1})\|^{2}_{2}\\
&-2\langle \mathbf{S_{1}}(\mathbf{u}^{n+1}-\mathbf{u}^{n}), \mathbf{u}_{e}^{n+1}\rangle
-2\langle \mathbf{S_{2}}(\mathbf{k}^{n+1}-\mathbf{k}^{n}), \mathbf{k}_{e}^{n+1}\rangle\\
\leq&(\beta\eta)^{-1}(\|\mathbf{\lambda}_{e}^{n}\|^{2}-\|\mathbf{\lambda}_{e}^{n+1}\|^{2})-(2-\eta)\beta\|\epsilon (\mathbf{u}^{n+1}, \mathbf{k}^{n+1)}\|^{2}\\
&+2\alpha_{n+1}-\|\mathbf{k}^{n+1}-\mathbf{k}^n\|^{2}_{\mathbf{S_{2}}}+\|\mathbf{k}^{n}-\mathbf{k}^{n-1}\|^{2}_{\mathbf{S_{2}}}\\
&-\beta\|\mathbf{k}^{n+1}-\mathbf{k}^n\|^{2}-\beta\|{\mathbf{k}_{e}^{n+1}}\|^{2}+\beta\|{\mathbf{k}_{e}^{n}}\|^{2}\\
&-\|\mathbf{u}^{n+1}-\mathbf{u}^n\|^{2}_{\mathbf{S_{1}}}-\|{\mathbf{u}_{e}^{n+1}}\|_{\mathbf{S_{1}}}^{2}+\|{\mathbf{u}_{e}^{n}}\|_{\mathbf{S_{1}}}^{2}\\
&-\|\mathbf{k}^{n+1}-\mathbf{k}^n\|^{2}_{\mathbf{S_{2}}}-\|{\mathbf{k}_{e}^{n+1}}\|_{\mathbf{S_{2}}}^{2}+\|{\mathbf{k}_{e}^{n}}\|_{\mathbf{S_{2}}}^{2}.}
\end{equation}
Define
\begin{equation}\label{14}
\cases{\delta_{n+1} = \min\{\eta, 1+\eta-\eta^{2}\}\beta\|\mathbf{k}^{n+1}-\mathbf{k}^n\|^{2}+\|\mathbf{k}^{n+1}-\mathbf{k}^n\|^{2}_{\mathbf{S_{2}}}, \\
t_{n+1} = \delta_{n+1}+\|\mathbf{u}^{n+1}-\mathbf{u}^n\|^{2}_{\mathbf{S_{1}}}+2\|\mathbf{u}^{n+1}-\bar{\mathbf{u}}\|_{\sum_{\mathbf{F}}}^{2}
+2\|\mathbf{k}^{n+1}-\bar{\mathbf{k}}\|_{\sum_{\mathbf{G}}}^{2}, \\
\psi_{n+1} = \theta(\mathbf{u}^{n+1}, \mathbf{k}^{n+1}, \mathbf{\lambda}^{n+1})+\|\mathbf{k}^{n+1}-\mathbf{k}^n\|^{2}_{\mathbf{S_{2}}}, \\
\theta(\mathbf{u}, \mathbf{k}, \mathbf{\lambda})=(\beta\eta)^{-1}\|\mathbf{\lambda}-\bar{\mathbf{\lambda}}\|^{2}+\|\mathbf{u}-\bar{\mathbf{u}}\|^{2}_{\mathbf{S_{1}}}+
\|\mathbf{k}-\bar{\mathbf{k}}\|^{2}_{\mathbf{S_{2}}}
+\beta\|\mathbf{k}-\bar{\mathbf{k}}\|^{2}.}
\end{equation}
Next,   we discuss two cases:\\
Case \uppercase\expandafter{\romannumeral1}: $\eta\in(0, 1]$.  It is obvious that
\begin{equation*}
2\langle \mathbf{k}^{n+1}-\mathbf{k}^{n}, \epsilon (\mathbf{u}^{n}, \mathbf{k}^{n})\rangle\leq\|\mathbf{k}^{n+1}-\mathbf{k}^n\|^{2}+\|\epsilon (\mathbf{u}^{n}, \mathbf{k}^n)\|^{2}.
\end{equation*}
By the definition of $\alpha_{n+1}$ and (\ref{13}),  we see
\begin{equation}\label{15}
\eqalign{&\psi_{n+1}+(1-\eta)\beta\|\epsilon(\mathbf{u}^{n+1}, \mathbf{k}^{n+1})\|^{2}-[\psi_{n}+(1-\eta)\beta\|\epsilon (\mathbf{u}^{n}, \mathbf{k}^n)\|^{2}]\\
&+t_{n+1}+\beta\|\epsilon(\mathbf{u}^{n+1}, \mathbf{k}^{n+1})\|^{2}\leq 0.\\}
\end{equation}
Case \uppercase\expandafter{\romannumeral2}: $\eta\in(1, \frac{1+\sqrt{5}}{2})$.
Similarly,  we have
\begin{equation}\label{16}
\eqalign{&\psi_{n+1}+(1-\eta^{-1})\beta\|\epsilon(\mathbf{u}^{n+1}, \mathbf{k}^{n+1})\|^{2}-[\psi_{n}+(1-\eta^{-1})\beta\|\epsilon (\mathbf{u}^{n}, \mathbf{k}^n)\|^{2}]\\
&+t_{n+1}+\eta^{-1}(1+\eta-\eta^{2})\beta\|\epsilon(\mathbf{u}^{n+1}, \mathbf{k}^{n+1})\|^{2}\leq0.\\}
\end{equation}
Denote
\begin{equation*}
g_{n+1}=\cases{\psi_{n+1}+(1-\eta)\beta\|\epsilon(\mathbf{u}^{n+1}, \mathbf{k}^{n+1})\|^{2}, & $\eta\in(0, 1];$\\
\psi_{n+1}+(1-\eta^{-1})\beta\|\epsilon(\mathbf{u}^{n+1}, \mathbf{k}^{n+1})\|^{2}, & $\eta\in (1, \frac{1+\sqrt{5}}{2}).$\\}
\end{equation*}
%when $\eta\in(0, 1]$,  $g_{n+1}=\psi_{n+1}+(1-\eta)\beta\|\epsilon(\mathbf{u}^{n+1}, \mathbf{k}^{n+1})\|^{2}$,
%when $\eta\in (1, \frac{1+\sqrt{5}}{2})$,
%$g_{n+1}=\psi_{n+1}+(1-\eta^{-1})\beta\|\epsilon(\mathbf{u}^{n+1}, \mathbf{k}^{n+1})\|^{2}$.\\
Thus,  from (\ref{15}) and (\ref{16}),  we conclude that the sequence $\{g_n\}$
is bounded and monotonically decreasing,
hence,  it has limits. Since $\psi_{n}>0$,  the sequence $\{\psi_{n}\}$ is bounded.

Let $\gamma=1$ or $\gamma=\eta^{-1}(1+\eta-\eta^{2})$. Furthermore,   from (\ref{15}) and (\ref{16}),
\begin{equation*}
0\leq t_{n+1}+\gamma\beta\|\epsilon(\mathbf{u}^{n+1}, \mathbf{k}^{n+1})\|^{2}\leq g_{n}-g_{n+1}.    
\end{equation*}
Thus,
\begin{equation}\label{17}
t_{n+1}\rightarrow0,
\end{equation}
\begin{equation}\label{18}
\|\epsilon(\mathbf{u}^{n+1}, \mathbf{k}^{n+1})\|\rightarrow0.
\end{equation}
Considering the relationship $\mathbf{\lambda}^{n+1}-\mathbf{\lambda}^{n}=(\eta\beta)\epsilon(\mathbf{u}^{n+1}, \mathbf{k}^{n+1})$,  we have
$\|\mathbf{\lambda}^{n+1}-\mathbf{\lambda}^{n}\|\rightarrow0$.
By (\ref{17}) and the bounded property of $\{{\psi_{n}}\}$,  one can get these sequences
$\{{\|\mathbf{\lambda}^{n+1}\|}\}$, $\{{\|\mathbf{u}^{n+1}_{e}}\|^{2}_{\mathbf{S_{1}}}\}$,
$\{{\|\mathbf{u}^{n+1}_{e}}\|^{2}_{\sum_{\mathbf{F}}}\}$,
$\{{\|\mathbf{k}^{n+1}_{e}}\|^{2}_{\sum_{\mathbf{G}}}\}$,
$\{{\|\mathbf{k}^{n+1}_{e}}\|^{2}_{\mathbf{S_{2}}}\}, \{{\|\mathbf{k}^{n+1}_{e}}\|^{2}\}$ are bounded.
By using the inequality
\begin{equation*}
\|\nabla\mathbf{u}^{n+1}_{e}\|\leq\|\nabla\mathbf{u}^{n+1}_{e}-\mathbf{k}^{n+1}_{e}\|+\|\mathbf{k}^{n+1}_{e}\|
=\|\epsilon(\mathbf{u}^{n+1}, \mathbf{k}^{n+1})\|+\|\mathbf{k}^{n+1}_{e}\|, 
\end{equation*}
we deduce that $\{\|\nabla\mathbf{u}^{n+1}_{e}\|\}$ is bounded,  so $\{\|\mathbf{u}^{n+1}_{e}\|_{\nabla^{\top}\nabla}\}$ is bounded.
Since $\|\mathbf{u}^{n+1}_{e}\|=\|\mathbf{u}^{n+1}_{e}\|_{S_{1}+\sum_{\mathbf{F}}+\nabla^{\top}\nabla+I}
-\|\mathbf{u}^{n+1}_{e}\|_{S_{1}+\sum_{\mathbf{F}+\nabla^{\top}\nabla}}$,  and the positive definite property of
$S_{1}+\sum_{\mathbf{F}}+\nabla^{\top}\nabla+I$,  the sequence $\{\|\mathbf{u}^{n+1}_{e}\|\}$ is bounded,  and hence the sequence $\{\|\mathbf{u}^{n+1}\|\}$ is bounded. Therefore,  the
sequence $\{(\mathbf{u}^{n}, \mathbf{k}^{n}, \mathbf{\lambda}^{n})\}$ is bounded,  which implies the existence of a convergent
subsequence to a clusters point,  denoted as
%$\{(\mathbf{u}^{n^{i}}, \mathbf{k}^{n^{i}}, \mathbf{\lambda}^{n^{i}})\}$£¬
$(\mathbf{u}^{n^{i}}, \mathbf{k}^{n^{i}}, \mathbf{\lambda}^{n^{i}})\rightarrow
(\mathbf{u}^{*}, \mathbf{k}^{*}, \mathbf{\lambda}^{*})$.

It follows from
(\ref{14}) and (\ref{17}) that
\begin{equation}\label{19}
\eqalign{ &\lim_{n\to \infty}\|\mathbf{k}^{n+1}-\mathbf{k}^{n}\|=0,\\
&\lim_{n\to \infty}\|\mathbf{k}^{n+1}-\mathbf{k}^{n}\|_{S_{2}}=0,\\
&\lim_{n\to \infty}\|\mathbf{u}^{n+1}-\mathbf{u}^{n}\|_{S_{1}}=0.}
\end{equation}
%\begin{equation}\label
%\cases{
%\lim_{n\to \infty}\|\mathbf{k}^{n+1}-\mathbf{k}^{n}\|=0, \\
%\lim_{n\to \infty}\|\mathbf{k}^{n+1}-\mathbf{k}^{n}\|_{S_{2}}=0, \\
%\lim_{n\to \infty}\|\mathbf{u}^{n+1}-\mathbf{u}^{n}\|_{S_{1}}=0.
%}
%\end{equation}
%\begin{equation}\label{19}
%\qquad \quad \fl \lim_{n\to \infty}\|\mathbf{k}^{n+1}-\mathbf{k}^{n}\|=0, \\
%\lim_{n\to \infty}\|\mathbf{k}^{n+1}-\mathbf{k}^{n}\|_{S_{2}}=0, \\
%\lim_{n\to \infty}\|\mathbf{u}^{n+1}-\mathbf{u}^{n}\|_{S_{1}}=0.
%\end{equation}
Thus,  from
\begin{equation*}
\eqalign{ \|\nabla \mathbf{u}^{n+1}-\mathbf{k}^{n}\|
%&=\|\nabla \mathbf{u}^{n+1}-\mathbf{k}^{n+1}
%+\mathbf{k}^{n+1}-\mathbf{k}^{n}\|\\
&\leq\|\nabla \mathbf{u}^{n+1}-\mathbf{k}^{n+1}\|+\|\mathbf{k}^{n+1}-\mathbf{k}^{n}\|,}
%&{\color{red}{(note:\|\nabla \mathbf{u}^{n+1}-\mathbf{k}^{n+1}\|=\epsilon(\mathbf{u}^{n+1}, \mathbf{k}^{n+1}))}},
\end{equation*}
we have $\lim_{n\to \infty}\|\nabla \mathbf{u}^{n+1}-\mathbf{k}^{n}\|=0$.
Taking limits on both sides of equation (\ref{6}) along the subsequence $(\mathbf{u}^{n^{i}}, \mathbf{k}^{n^{i}}, \mathbf{\lambda}^{n^{i}})$ and using  the closedness of subdifferential,  we have
\begin{equation*}
\cases{-\nabla^{\top}\mathbf{\lambda}^{*}=\nabla F (\mathbf{u}^{*}), \\
\mathbf{\lambda}^{*}\in\partial G (\mathbf{k}^{*}), \\
0=\nabla \mathbf{u}^{*}-\mathbf{k}^{*}.}
\end{equation*}
Thus,  $(\mathbf{u}^{*}, \mathbf{k}^{*})$ is the optimal solution of (\ref{yue}) and $\mathbf{\lambda}^{*}$ is the corresponding Lagrange multiplier.

Now,  let's prove that ${\displaystyle\lim_{n\rightarrow\infty}}(\mathbf{u}^{n}, \mathbf{k}^{n}, \mathbf{\lambda}^{n})=(\mathbf{u}^{*}, \mathbf{k}^{*}, {\lambda}^{*})$.
Since $(\mathbf{u}^{*}, \mathbf{k}^{*}, {\lambda}^{*})$ satisfies (\ref{6}),  we could replace $(\bar{\mathbf{u}}, \bar{\mathbf{k}}, \bar{\mathbf{\lambda}})$ with $(\mathbf{u}^{*}, \mathbf{k}^{*}, \mathbf{\lambda}^{*})$ in the above analysis. From (\ref{15}),  (\ref{16}),  (\ref{17}) and (\ref{18}),  we find $g_{n_{i}}\rightarrow0, n_{i}\rightarrow\infty$.
(\ref{15}) and (\ref{16}) indicate the sequence $\{g_{n}\}$ has limits,  then $g_{n}\rightarrow0, n\rightarrow\infty$. So $\psi_{n}\rightarrow0, n\rightarrow\infty$. Further,  we know by the definition of $\psi_{n}$ that
\begin{equation}\label{21-2}
\eqalign{ &\|\mathbf{\lambda}^{n}-\mathbf{\lambda}^{*}\|\rightarrow0\Longrightarrow\mathbf{\lambda}^{n}\rightarrow\mathbf{\lambda}^{*},\\
&\|\mathbf{k}^{n}-\mathbf{k}^{*}\|\rightarrow0\Longrightarrow\mathbf{k}^{n}\rightarrow\mathbf{k}^{*},\\
&{\|\mathbf{u}^{n+1}_{e}}\|^{2}_{\mathbf{S_{1}}}\rightarrow0.}
\end{equation}
Considering (\ref{17}) and (\ref{18}),  we get
\begin{equation}\label{22}
{\|\mathbf{u}^{n+1}_{e}}\|^{2}_{\mathbf{\sum_{F}}}\rightarrow0,  n\rightarrow\infty.
\end{equation}
Then,  by (\ref{21-2}) and (\ref{22}),
\begin{equation}\label{23}
{\|\mathbf{u}^{n}_{e}}\|_{\mathbf{S_{1}+\sum_{F}}}\rightarrow0, n\rightarrow\infty.
\end{equation}
From the positive semidefinite property of $\mathbf{\sum_{F}}$ and the positive definite property of $\mathbf{S_{1}}$,  it is clear that
\begin{equation*}
\|\mathbf{u}^{n}_{e}\|^{2}=\langle\mathbf{u}^{n}_{e}, \mathbf{u}^{n}_{e}\rangle\leq\|\mathbf{u}^{n}_{e}\|_{\mathbf{S_{1}+\sum_{F}}}\|\mathbf{u}^{n}_{e}\|_{(\mathbf{S_{1}+\sum_{F}})^{-1}},     
\end{equation*}
which combined with (\ref{23}) gives ${\displaystyle\lim_{n\rightarrow\infty}}\|\mathbf{u}^{n}_{e}\|=0$,  namely, ${\displaystyle\lim_{n\rightarrow\infty}}\mathbf{u}^{n}=\mathbf{u}^{*}$.

In brief,  when $\eta\in(0, \frac{1+\sqrt{5}}{2})$,  we have
\begin{equation*}
\lim_{n\rightarrow\infty}(\mathbf{u}^{n}, \mathbf{k}^{n}, \mathbf{\lambda}^{n})=(\mathbf{u}^{*}, \mathbf{k}^{*}, {\lambda}^{*}).    
\end{equation*}

\section*{References}
\bibliographystyle{plain}
\bibliography{01-iopart}

\end{document}